\documentclass[useAMS,usenatbib]{mn2e}

\usepackage{amssymb}
\usepackage{amsmath}
\usepackage{multirow}
\usepackage{epsfig}

\providecommand{\eg}{e.g.}
\providecommand{\etal}{et~al.}

\def\la{\mathrel{\hbox{\rlap{\lower.55ex \hbox {$\sim$}}
        \kern-.3em \raise.4ex \hbox{$<$}}}}
\def\ga{\mathrel{\hbox{\rlap{\lower.55ex \hbox {$\sim$}}
        \kern-.3em \raise.4ex \hbox{$>$}}}}

\def\swift{{\sl Swift\/}}

\def\AA{\em A.\& A.}

\def\APJ{\em ApJ.}
\def\APJl{\em ApJ.Lett}

\def\APS{\em Astrophys. \& Space Sci.}

\def\AST{\em Astron. J.}

\def\MRA{\em MNRAS}
\def\NAT{\em Nature}

\def\PAS{\em PASP}

\def\be{\begin{equation}}
\def\ee{\end{equation}}
\def\bea{\begin{eqnarray}}
\def\eea{\end{eqnarray}}
\def\bes{\begin{equation*}}
\def\ees{\end{equation*}}
\def\beas{\begin{eqnarray*}}
\def\eeas{\end{eqnarray*}}

\title{GRB~060607A: A GRB with Bright Asynchronous Early $X$-ray and Optical 
Emissions}

\author[Ziaeepour \& \etal]{Houri~Ziaeepour$^{1}$
        \thanks{Email: hz@mssl.ucl.ac.uk},
        Stephen~T.~Holland$^{2,3,4}$,
        Patricia~T.~Boyd$^{5}$,
        Kim~Page$^{6}$, \and
        Samantha~Oates$^{1}$,  
        Craig~B.~Markwardt$^{5}$,
        Peter~M{\'e}sz{\'a}ros$^{7}$,
        Neil~Gehrels$^{2}$, \and 
        Francis~E.~Marshall$^{4}$,
        Jay~Cummings$^{4}$,
        Mike~Goad$^{6}$\\
$^{1}$Mullard Space Science Laboratory, Holmbury St Mary, Dorking, 
Surrey RH5 6NT, UK\\
$^{2}$Astrophysics Science Division, NASA Goddard Space 
Flight Centre, Greenbelt, MD 20771, USA\\
$^{3}$Universities Space Research Association, Suite 620, 10211 Winicopin 
Circle, Columbia, MD 21044--3431, USA\\
$^{4}$CRESST, Room 260, Building 1, NASA Goddard Space Flight Centre, 
Greenbelt, MD 20771, USA\\
$^{5}$Laboratory for Astroparticle Physics, NASA Goddard Space Flight Centre, 
Greenbelt, MD 20771, USA\\
$^{6}$Department of Physics and Astronomy, University of Leicester,Leicester, LE1 7RH, UK\\
$^{7}$Dept.\ of Astronomy and Astrophysics, Pennsylvania State University, 
525 Davey Laboratory, University Park, PA 16802, USA}

\begin{document}

\date{Accepted $\ldots$; Received $\ldots$; in original form Oct 2007}

\pagerange{\pageref{firstpage}--\pageref{lastpage}} \pubyear{2007}

\maketitle

\label{firstpage}

\begin {abstract} 
The early optical emission of the moderately high redshift ($z=3.08$) 
GRB 060607A shows a remarkable broad and strong peak with a rapid rise and a 
relatively slow power-law decay. It is not coincident with the strong 
early-time flares seen in the X-ray and gamma-ray energy bands. There is 
weak evidence for variability superposed on this dominant component 
in several optical bands that can be related to flares in high energy bands. 
While for a small number of GRBs, well-sampled optical flares have been 
observed simultaneously with X-ray and gamma ray pulses, GRB 060607A is one 
of the few cases where the early optical emission shows no significant 
evidence for correlation with the prompt emission. In this work we first 
report 
in detail the broad band observations of this burst by {\sl Swift\/}. Then 
by applying a simple model for the dynamics and the synchrotron radiation 
of a relativistic shock, we show that the dominant component of the early 
emissions in optical wavelengths has the same origin as the tail emission 
produced after the main gamma ray activity. The most plausible explanation 
for the peak in the optical light curve seems to be the cooling of the 
prompt after the main collisions, shifting the characteristic synchrotron 
frequency to the optical bands. The fact that the early emission in X-ray 
does not show a steep decay, like what is observed in many other 
GRBs, is further evidence for slow cooling of the prompt shell within this 
GRB. It seems that the cooling process requires a steepening of the 
electron energy distribution and/or a break in this distribution at high 
energies. From simultaneous gamma-ray emission during the first flare, 
the behaviour of hardness ratio, and the lack of spectral features, we 
conclude that the X-ray flares are due to the collision of late shells 
rather than late reprocessing of the central engine activities. The sharp 
break in the X-ray light curve at few thousands of seconds after the trigger, 
is not observed in the IR/optical/UV bands, and therefore can not be a 
jet break. Either the X-ray break is due to a change in the spectrum of 
the accelerated electrons or the lack of an optical break is due to the 
presence of a related delayed response component.

\end {abstract}




\section{Introduction}\label{SECTION:intro}
GRB 060607A was a long, fairly hard GRB localized by the {\sl Swift\/} Burst 
Alert Telescope (BAT)\citep{B2005,bar05} at $T_0 \simeq$ 05:12:13 UT on 2006 
June 7\citep{HZGCN}. The {\sl Swift\/} spacecraft rapidly slewed, directing 
the X-Ray Telescope (XRT)\citep{BET2005} and the Ultraviolet and Optical 
Telescope (UVOT)\citep{RO2005} at the BAT position. The observations 
commenced at $T_0 + 63.6$~sec after the BAT trigger in the optical/UV and 
$T_0 + 73.6$~sec in the X-ray, providing broad-band spectral and high 
time resolution light curves in the X-ray, optical, and ultraviolet. The X-ray 
light curve shows two bright flares, the first of which at $T_0 + 98$ sec is 
also observed in the BAT energy range $15-300$~keV. The second flare occurs 
at $T_0 + 260$ sec and is marginally (if at all) detected by BAT (see 
Fig. \ref{FIGURE:lightcurves}). The optical emission is at first quite faint, 
then rises nearly three magnitudes over a time span of about $100$ sec. The 
peak in the optical is not simultaneous with the X-ray flares.

Historically, ground-based follow-up optical observations of GRBs usually 
started at least several minutes after the burst occurred, and it has not been 
unusual for optical observations to start several hours into the afterglow 
phase. Over the past few years fast-slewing robotic telescopes e.g. 
\citep{B2001,PMR2001,VBB2002,AKM2003,CSS2004} and the {\sl Swift\/} mission 
\citep{GCG2004} have made it possible to observe the first few minutes of 
optical emission from GRBs. In some cases: 
GRB 990123 \citep{ABB1999}, GRB 041219A \citep{VWW2005}, GRB 050820A 
\citep{VWW2006}, GRB 051109A, GRB 051111\citep{Y2007}, GRB 060124 
\citep{RCC2006}, GRB 060418 \citep{MVM2006}, GRB 060526 \citep{D2007}, 
GRB 060607A, GRB 061007 \citep{S2006}, GRB 061121 \citep{P2007}, GRB 070616 
\citep{STG2007,ST2007} the optical emission has been observed during the 
prompt 
gamma-ray emission. In the case of GRB 990123 there is no correlation between 
the structure of the gamma-ray light curve and the optical flux 
\citep{ABB1999}. If anything, there is an apparent anti-correlation, although 
the optical data is not well sampled. On the other hand, all the other 
bursts above show a correlation between the optical and gamma-ray fluxes. 
In the case of GRB 060124, although the optical flux slightly increases 
during the main peak at about $600$~sec, its relative rise is much smaller 
than the rise in gamma-ray and X-ray fluxes. In the case of GRB 060607A, 
clearly there is no correlation between the 
dominant component of the early optical and X-ray emissions. The 
optical flux arrived at its peak when the X-ray emission was decreasing. A 
similar behaviour was also observed in GRB 060418 \citep {MVM2006}.

It is not yet clear how the prompt gamma-ray emission is related to the later 
emissions in the lower energy bands. {\sl Swift\/} XRT observations have 
shown that the X-ray early emission of the most GRBs exhibit a very steep 
power-law decline with ($3 \lesssim \alpha \lesssim 5$, where the flux at 
time $t$ is $f_\nu \propto t^{-\alpha}$) decay within 
$T_0 + \lesssim 1000$~sec \citep{NKG2006,OB2006,BUT2006,WI2007}). This decay 
was expected to be caused by the high-latitude emission from the internal 
shocks that are driving the prompt high-energy emission, and therefore 
is not produced by the same 
mechanism that drives the late-time afterglow \citep[{\eg}][]{ZK2005,LZO2006}, 
presumably the external shock with the ISM or wind. However, this rapid 
decay phase has not been observed in any optical early emission. Moreover, 
the high-latitude emission should satisfy a strict relation between 
$\alpha$ and the photon index $\beta$: $\alpha = 2 + \beta$ \citep{Fen1996}. 
This relation is however only satisfied in a small fraction of GRBs with 
a steeply decaying X-ray tail emission which also do not show significant 
spectral evolution \citep{ZLA2007}. The spectral evolution of the tail 
emission is another argument against high latitude origin of the tail. 
Therefore, one can conclude that the high-latitude emission is not a 
dominant contributor in the tail emission. The fact that the light curves 
in different bands in general 
do not follow each other suggests that either multiple components should be 
involved \citep{KMP2006} and/or processes are chromatic and the emission 
evolves both in time and in energy. Therefore, one should expect various 
degrees of correlation between the light curves in the different energy bands 
depending on the internal properties of the system and its environment. 
GRB 060607A is a good example of how peculiar the relation between 
energy bands can be. The initial decay slope in X-ray is shallower than most 
bursts $\alpha \sim 1$, and the optical light curve shows a peak uncorrelated 
to the flares in gamma-ray and X-ray.

In this paper we focus on the early optical and X-ray emissions as observed by 
{\sl Swift\/}'s UVOT and XRT instruments. The goal is to compare the 
multi-band {\sl Swift\/} data, with a simple relativistic shock and 
synchrotron emission model, and to try to reconstruct the 
history of events leading to the prompt and afterglow emission as observed by 
{\sl Swift\/}. In Section \ref{SECTION:obs}, we describe the {\sl Swift\/} 
observations. In Section \ref{SECTION:timevar}, we present the time variation 
analysis, and in Section \ref{SECTION:specvar}, we discuss broad-band spectral 
variability during the early emission phase and give a qualitative 
interpretation of the data. In Section \ref{SECTION:modelling}, we apply 
the model mentioned above to the data to interpret observations and to 
estimate some of the parameters. Finally, we summarize our results in 
Section \ref{SECTION:conclu}. The theoretical model used in Section 
\ref{SECTION:modelling} is reviewed briefly in Appendix \ref{appendixa}. 
In Appendix \ref{appendixb} the conditions determining 
the reliability of the extrapolation of BAT light curve to the XRT energy band 
is discussed.

\section{Observations}\label{SECTION:obs}
In this section we briefly report the result of the analysis of the BAT, 
XRT, and UVOT data. Due to the peculiarity of the optical afterglow  
of this burst and its importance for the interpretation of the GRB, the 
UVOT data is discussed with more details.

\subsection{BAT Data}\label{SECTION:bat}
\begin{figure}
\includegraphics[height=7cm]{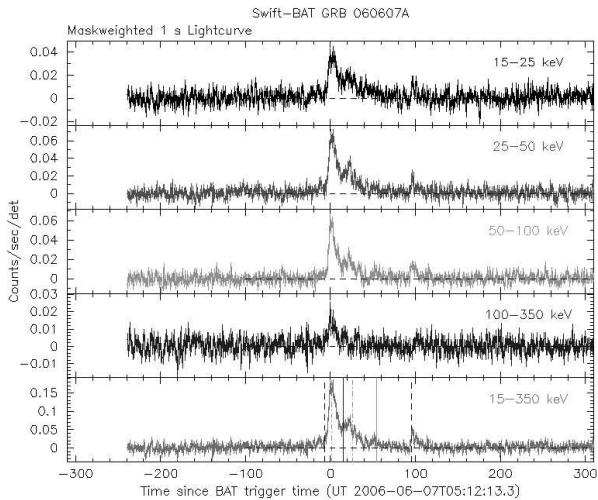} 
\caption{The mask
weighted light curves, at $1$~sec time resolution, for the BAT energy
bands. The double FRED-like structure is visible at the start of all
of the BAT light curves and a late peak is seen at approximately $100$~sec.
Vertical lines: $T_90$ interval (dash lines), $T_50$ interval (dash-dot 
lines), slew interval (full lines).
\label{FIGURE:bat_lc}}
\end{figure}

The BAT light curve showed a double-peaked structure with a duration of 
about 40 sec \citep{TBB2006}. The peak count rate was approximately 
$3000$ count sec$^{-1}$ ($15-350$~keV) at the time of the trigger. The 
mask-weighted light curve (Fig. \ref{FIGURE:bat_lc}) consists of two 
overlapping FRED-like peaks from $T_0 - 5$~sec to $T_0 + 40$~sec. There is 
a second double peaked structure between $T_0 + 95$~sec and $T_0 + 105$~sec. 
$T_{90}$ ($15-350$~keV) is $100 \pm 5$~sec (estimated error including 
systematics).

The time-averaged spectrum from $T_0 - 14.1$~sec to $T_0 + 104.5$~sec is best 
fit by a simple power-law model with a power-law index of 
$\Gamma  = 1.45 \pm 0.07$. The fluence in the $15-150$~keV band is 
$2.6 \pm 0.1 \times 10^{-6}$~erg~cm$^{-2}$. The $1$~sec peak 
photon flux measured from $T_0 - 0.97$~sec in the $15-150$~keV band is 
$1.4 \pm 0.1$~ph~cm$^{-2}$~sec$^{-1}$. All the quoted errors are at the 
$90\%$ confidence level.

Using the spectroscopic redshift of this burst $z = 3.082$ reported by 
\citep{LVS2006} and a cosmology with $\Omega_m = 0.3$, 
$\Omega_\Lambda  = 0.7$, 
and $H_0 = 65$, we find $E_{iso}$ ($1-1000$~keV in the rest frame) to be 
$1.1 \times 10^{53}$~erg. This is based on an extrapolation of the BAT 
power-law fit into the corresponding observer energy band.

\subsection {XRT Data}\label{SECTION:xrt}
Observations using the {\sl Swift\/} X-Ray Telescope (XRT) began 
$73.6$~sec after the trigger. It found a bright, variable, uncatalogued 
X-ray source \citep{P2006}. The XRT position enhanced by UVOT astrometry 
\citep {G2007} was at RA(J2000) = 21:58:50.46, Dec(J2000) = -22:29:47.3, with 
an estimated uncertainty of $1.6''$ ($90\%$ confidence radius).

\begin{figure}
\includegraphics[height=6cm,angle=-90]{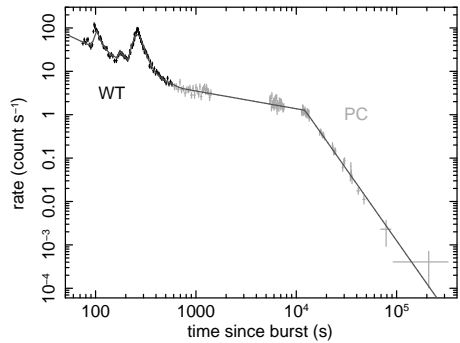}
\caption[GRB060607-xrt_lc.eps]{The XRT $X$-ray light
curve (0.3-10 keV), WT (black), PC (grey). This plot includes all the XRT 
observations of GRB 060607A.
\label{FIGURE:xrt_lc}}
\end{figure}

The XRT light curve shows three flares peaking at approximately $97$~sec, 
$175$~sec, and $263$~sec after the BAT trigger. These flares are superposed 
on a decaying continuum with a decay index $\alpha_1 = 1.09 \pm 0.04$. 
At $714^{+88}_{-94}$~sec after the BAT trigger the slope flattens, becoming 
$\alpha_2 = 0.41 \pm 0.03$. There is a second break at 
$12200^{+360}_{-350}$~sec after which the decay becomes 
$\alpha_3 = 3.29^{+0.11}_{-0.1}$. The X-ray light curve of GRB 060607A is 
shown in Fig. \ref{FIGURE:xrt_lc}. 

\subsection {UVOT Data}\label{SECTION:uvot}
The {\sl Swift\/} Ultraviolet and Optical Telescope (UVOT) observations began 
$63.6$~sec after the trigger 
with a $10$~sec settling mode exposure with the V filter, followed by finding 
chart exposures of $100$~sec with the White ($160-650$~nm) filter starting at 
$T_0 + 73.6$~sec and then $400$~sec $V$ filter \citep{O2006}. The settling 
exposure and each of the two finding chart exposures are taken in 
"Image \& Event" mode, so in addition to the full-frame integrated image, 
event-by-event data is also available at $11.0322$~msec maximum time 
resolution.

\begin{figure}
\includegraphics[height=8cm]{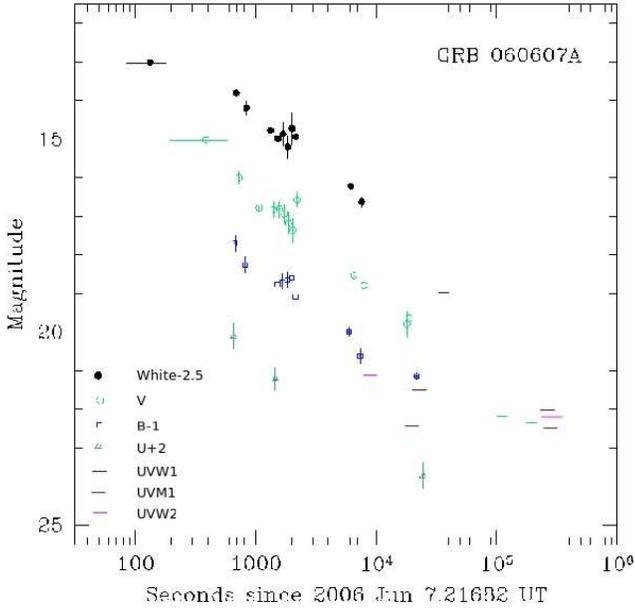} \caption{UVOT light
curves in White, $V$, $B$, and $U$ extending out to where only 3-sigma
upper limits are seen. They are calculated from co-added exposures. The 
background was measured in annulus around the source position as explained 
in the text. The first imaging exposure from $\sim 73$~sec. to $\sim 172$~sec
includes the rising part of the light curve, and therefore the rise of the 
optical flux is only observed in event-by-event data of the UVOT (see 
Fig.\ref{FIGURE:lightcurves}). Note also 
the episode of flattening, or possible rebrightening, between $\sim 1000$~sec 
and $\sim 2000$~sec.\label{FIGURE:uvot_data}}
\end{figure}

\begin{figure}
\includegraphics[height=8cm]{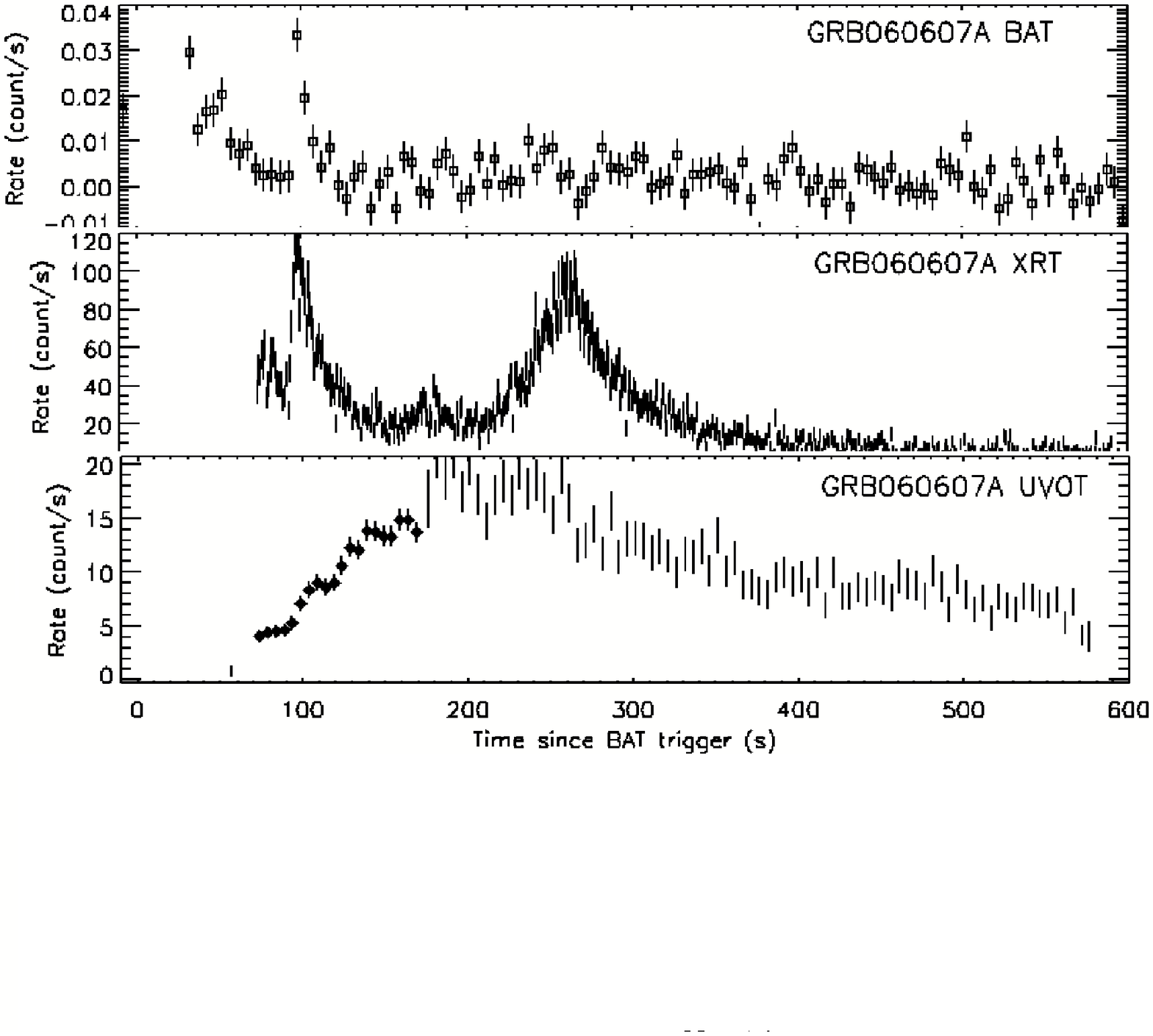} 
\vspace{-2cm}\caption{The light curves, at $5$~sec 
time resolution, for the BAT (upper), XRT (middle), and UVOT (bottom). 
The prompt double FRED-like structure as well as later spike at $\sim 100$~sec 
are visible in the BAT light curve. The XRT light curve shows probably a 
minor at the beginning of observations following by the flare at 
$\sim 100$~sec detected also 
by the BAT, then one minor flare at $\sim 170$~sec, and one major flare at 
$\sim 260$~sec. In the UVOT light curve 
circles correspond to the White filter and bars to V filter data points. 
The UVOT V band count rates are weighted by a factor of $\sim3$. 
The optical light curve shows some variations at the time of flares, 
otherwise it is dominated by a continuous component peaking at $\sim 200$~sec.
\label{FIGURE:lightcurves}}
\end{figure}

A bright afterglow was detected in autonomous ground processing software, 
at RA = 21:58:50.40, Dec = -22:29:46.7 (J2000) with a $1\sigma$ error 
radius of approximately $0.5''$. This position was $4.7''$ from the center 
of the refined XRT error circle and is $0.95''$ from the XRT position 
enhanced by UVOT astrometry. The estimated initial White magnitude in the 
finding chart image was $15.7$ with a $1\sigma$ error of $\approx 0.5$~mag. 
The optical afterglow was also detected in White, $V$, 
$B$, and $U$ filters (Fig.\ref{FIGURE:uvot_data}). The non-detections in the 
ultraviolet (UV) bands are consistent with the spectroscopic redshift of 
$z = 3.082$ \citep{LVS2006}. We have also extracted $5$~sec binned light 
curves from event-by-event data to investigate the short time variability 
of the UVOT afterglow and to compare it to the XRT and BAT light curves 
(Fig.\ref{FIGURE:lightcurves}). Throughout this work, all the investigations 
of the UVOT data up to $\sim 600$~sec is based on this light curve.

We performed photometry on each UVOT exposure using a circular aperture with 
a radius of $2''$ centred on the position of the optical afterglow. This 
radius is approximately equal to the full-width at half-maximum (FWHM) of 
the UVOT point-spread function (PSF). The PSF varies with filter and with 
the temperature of the telescope, so we did not match the extraction aperture 
to the PSF for each exposure. The PSF FWHM, averaged over the 
temperature variations, ranges from $1.''79 \pm 0.05''$ for the $V$ filter to 
$2.17'' \pm 0.03''$ for the UVW2 filter. The background was measured in a 
sky annulus of inner radius $17.5''$ and width $5''$ centred on the afterglow.

Aperture corrections were computed for each exposure to convert the $2''$ 
photometry to the standard aperture radii used to define UVOT's photometric 
zero points ($6''$ for UBV and $12''$ for the ultraviolet filters). Six 
isolated stars were used to compute the aperture correction for each 
exposure. The RMS scatter in the mean aperture correction for a single 
exposure was typically $\approx 0.02$ mag. The RMS scatter for each exposure 
was added in quadrature to the statistical error in the $2''$ magnitude to 
obtain the total 1$\sigma$ error in each point.

Since the UVOT is a photon-counting device it is only able to record one 
photon per detector cell during each read out. This results in coincidence 
losses at high count rates. For very high count rates, corresponding to 
$V \lesssim 13.5$, these losses are significant and can dramatically affect 
the photometry, so coincidence loss corrections must be made. We have 
corrected all of our data for coincidence loss, although for most 
of the observations of the afterglow this correction is negligible.

The values quoted in Table \ref{TABLE:phot} are not corrected for the expected 
Galactic reddening of $E_{B-V} = 0.03 \pm 0.02$ mag \citep{SFD1998}. This 
reddening corresponds to Galactic extinctions of $A_U = 0.15$, $A_B = 0.13$, 
and $A_V = 0.10$. The Galactic extinctions in the UVOT ultraviolet filters 
were calculated using the Milky Way extinction law from \citep {P1992}. The 
ultraviolet extinctions are $A_{UVW1} = 0.23$, $A_{UVM2} = 0.29$, and 
$A_{UVW2} = 0.21$ and $A_{White} = 0.13$.

The afterglow was detected by UVOT from $63.6$~sec after the BAT trigger until 
it faded below detectability at approximately $2 \times 10^4$~sec 
post-trigger. All significant detections as well as $3\sigma$ upper limits 
are listed in Table \ref{TABLE:phot} and shown in Fig. \ref{FIGURE:uvot_data}.

\begin{table}
\caption{{\sl Swift\/} UVOT Photometry of GRB 060607A\label{TABLE:phot}}
\begin{center}
\begin{tabular}{rcrcc}
\hline \\
Time + $T_0$ (sec) & Filter & Exposure (sec) & Magnitude & Error \\
\hline \\
         385 &        $V$ &        387 &      15.03  &         0.04 \\
         728 &        $V$ &         19 &      16.00  &         0.16 \\
        1065 &        $V$ &        392 &      16.78  &         0.05 \\
        1413 &        $V$ &         19 &      16.83  &         0.22 \\
        1571 &        $V$ &         19 &      16.81  &         0.21 \\
        1724 &        $V$ &         19 &      16.95  &         0.23 \\
        1883 &        $V$ &         19 &      17.18  &         0.27 \\
        2041 &        $V$ &         19 &      17.36  &         0.29 \\
        2199 &        $V$ &         19 &      16.58  &         0.20 \\
        6555 &        $V$ &        195 &      18.53  &         0.17 \\
        7947 &        $V$ &        127 &      18.79  &         0.36 \\
     18\,360 &        $V$ &        871 &      19.90  &         0.22 \\
    110\,400 &        $V$ &    15\,610 &    $>22.17$ &           - \\
    197\,400 &        $V$ &    22\,490 &    $>22.35$ &           - \\

         675 &        $B$ &         10 &      16.70  &         0.20 \\
         817 &        $B$ &         10 &      17.26  &         0.20 \\
        1513 &        $B$ &         19 &      17.77  &         0.20 \\
        1666 &        $B$ &         19 &      17.69  &         0.19 \\
        1825 &        $B$ &         19 &      17.65  &         0.20 \\
        1983 &        $B$ &         19 &      17.59  &         0.17 \\
        2136 &        $B$ &         19 &      18.08  &         0.28 \\
        5943 &        $B$ &        195 &      18.99  &         0.12 \\
        7372 &        $B$ &        193 &      19.62  &         0.20 \\
     21\,710 &        $B$ &       2663 &      20.13  &         0.08 \\

         654 &        $U$ &         19 &      18.11  &         0.33 \\
        1455 &        $U$ &        117 &      19.21  &         0.28 \\
     24\,580 &        $U$ &       5273 &      21.72  &         0.32 \\
    284\,300 &        $U$ &       1808 &    $>21.65$ &           - \\

         132 &      White &         97 &      15.52  &         0.06 \\
         686 &      White &         10 &      16.31  &         0.24 \\
         833 &      White &         10 &      16.70  &         0.18 \\
        1324 &      White &         94 &      17.28  &         0.08 \\
        1529 &      White &         10 &      17.49  &         0.20 \\
        1682 &      White &         10 &      17.37  &         0.29 \\
        1840 &      White &         10 &      17.71  &         0.29 \\
        1999 &      White &         10 &      17.23  &         0.41 \\
        2152 &      White &         10 &      17.45  &         0.51 \\
        6149 &      White &        188 &      18.72  &         0.10 \\
        7578 &      White &        195 &      19.13  &         0.13 \\
     36\,560 &      White &        747 &    $>21.48$ &          - \\

     24\,180 &       UVW1 &       4925 &    $>22.06$ &          - \\
    284\,000 &       UVW1 &       3376 &    $>22.00$ &          - \\

     23\,710 &       UVM2 &       3742 &    $>22.26$ &          - \\
    283\,600 &       UVM2 &       5539 &    $>22.64$ &          - \\

        9302 &       UVW2 &       1423 &    $>21.16$ &          - \\
    284\,700 &       UVW2 &       7086 &    $>22.16$ &          - \\
\hline
\end{tabular}
\end{center}
\footnote*{The first column shows the instant corresponding to 
the middle of the exposure since trigger time.}
\end{table}

There is not a unique way to fit light curves specially when they do not have 
a simple power-law behaviour. For the PROMPT telescope observations of 
GRB 060607A\citep{NH2006, NH2007} authors have used a complex expression 
including absorption terms and power-law rising and falling terms around 
each of the features in the light curve. Here we are mainly interested in 
the origin of the dominant component of the optical light curves. Therefore, 
a simple power-law separately fit on the rising and declining segments 
of the optical light curves is adequate. In Sec.\ref{SECTION:timevar} we 
also fit the optical light curve by adding a component proportional to the 
prompt gamma-ray to investigate the contribution of the prompt to optical 
emission. 

For the rising section of the optical emission from $\sim 100$~sec to 
$\sim 170$~sec only White filter observations are available. The light curve 
is fit by a rising power-law with $\alpha= 2.3 \pm 0.3$. The relatively large 
uncertainty is due to the varying components during flares in this time 
interval. The decay slopes are listed in Table \ref{TABLE:uvot_decay}. 
The weighted mean decay index is $\alpha= 0.9 \pm 0.06$ ($1\sigma$ error) 
for all times after 
$\approx 600$~sec after the BAT trigger. It is consistent with what 
is seen in other optical afterglows before the jet break occurs (Oates, 
{\etal} in preparation). 

The decay slope is the same before and after 
the fluctuation in the light curve between $1000$~sec and $2000$~sec. This 
feature was also observed by the PROMPT telescope \citep{NH2007} and 
by the REM telescope \citep{MVM2006}, and is probably the optical 
emission from a weak flare at about $1000$~sec after trigger, see Fig. 
\ref{FIGURE:xrt_lc}. Another feature was observed by the PROMPT telescope 
between $\sim 3000$ and $\sim 4000$~sec. However, a gap in the {\sl Swift\/} 
data from $\sim 1500$~sec to $5000$~sec does not allow to see if it is 
related to an X-ray flare.

There is also a faint feature in the UVOT light curve that at first sight 
can be considered as noise: A double peak at the time of maximum flux. 
This feature is independently observed also by the PROMPT \citep{NH2007} 
in $B$ and by the REM \citep{MVM2006} telescopes in $H$ filter, and 
therefore most probably is real. It can be related to the flare at 
$T_0 + 179$~sec. 

\begin{table}
\caption{{\sl Swift\/} UVOT Decay Slopes of GRB~060607A\label{TABLE:uvot_decay}}
\begin{center}
\begin{tabular}{llll}
\hline \\
Filter & $\alpha$ & Error ($1\sigma$) & $\chi^2/$(d.o.f) \\
\hline \\
    $V$   & 1.21 & 0.03 & 72.29/11 \\
    $B$   & 0.88 & 0.04 & 12.47/8  \\
    $U$   & 0.88 & 0.11 &  0.61/1  \\
   White  & 0.77 & 0.02 & 17.68/9  \\
\hline
\end{tabular}
\end{center}
\end{table}

\section{Analysis of Time Variability}\label{SECTION:timevar}
The brightness of the early X-ray and optical emissions of GRB 060607A 
together 
with good time resolution of the {\sl Swift\/} on-board instruments permit the 
investigation of the relation between various features observed in different 
energy bands. This is very important for identifying the related physical 
processes and their modeling. In this section we investigate the correlation 
between features.

As mentioned in the previous sections, it is very clear that a fast varying 
component is superposed on the dominant continuous component of the optical 
emission. In order to determine the contribution of this fast varying 
component to the optical afterglow during early UVOT observations, we 
performed an analysis 
similar to that of Vestrand \etal (2005). We assumed that the UVOT light curve 
could be represented by a component proportional to the gamma-ray component 
in the same time interval:
\begin{equation}
F_p (t) = C_p F_\gamma (t) \label {promptf} 
\end{equation}
and a continuous component of the form:
\begin{equation}
F_a (t) = C_a \biggl (\frac {t - t_0}{t_0}\biggr )^{-s} \exp\biggl 
(\frac {-\tau}{t - t_0} \biggr ) \Theta (t-t_0) \label {agf}
\end{equation}
where $t_0$ is an arbitrary initial time, $\tau$ is the timescale for the rise 
of the 
optical emission, $s$ is the power-law decay index, and $C_p$ is the ratio of 
the UVOT fast varying component in White filter (the only filter with 
simultaneous BAT detection) to the BAT $15-150$~keV flux, and $C_a$ is the 
amplitude of the continuous component of the optical emission. The step 
function $\Theta$ is added to restrict this equation to $t \geqslant t_0$.

\begin{figure}
\includegraphics[height=5cm]{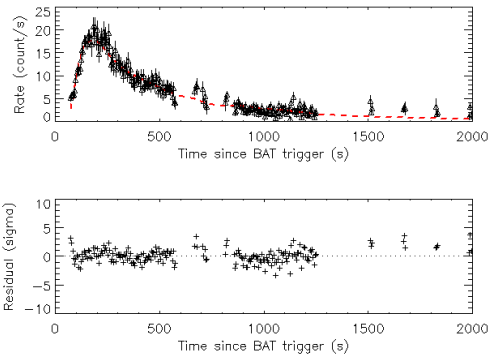} \caption[uvot_flare.eps]{The UVOT
White and $V$-band data are fit by the two-component model (dashed
line) described in Sect.~\ref{SECTION:timevar}.  The data from the two
filters have been arbitrarily shifted so that they are
aligned.\label{FIGURE:uvot_flare}}
\end{figure}

Our best fit is shown in Fig. \ref{FIGURE:uvot_flare}. The best fit has 
$t_0 = 37.9 \pm 2.3$~sec, $\tau = 209 \pm 5$~sec. The power-law index has 
been fixed to $s = 2$. The ratio of the fast varying component to the 
continuous afterglow in the UVOT data is $C_p/C_a = (6.25 \pm 8.20) \times 
10^{-2}$. This suggests that the fast varying optical emission makes a small, 
if not negligible, contribution to the observed optical light at the time of 
the flares. This is in contrast to what was found for some of other GRBs with 
simultaneous gamma ray, X-ray, and optical observations: 
GRB 041219A\citep{VBB2002}, GRB 050820A\citep{VWW2005}, 
GRB 060526\citep{D2007}, GRB 061121\citep{P2007}, where the prompt/flare 
component of the optical light makes a significant contribution to the total 
optical light. The possible reason can be the fact that for these bursts BAT 
had triggered on a faint precursor that produced a faint continuous emission 
before the occurrence of the main gamma-ray peak. In the case of GRB 060607A 
the main peak was at trigger time and had a significant tail emission. The 
time varying component here is due to fainter flares at later time, and 
therefore less significant than the remnant of the prompt emission.

To see whether the X-ray and the optical light curves correlate or more 
precisely if there is any trace of the flares in the UVOT light curve, we 
removed the smooth component of the rising section in White filter using the 
Savitzky-Golay smoothing algorithm \citep{SG1964}, and correlated the residue 
with the XRT light curve in $0.3-10$~keV. The feature in the rising part of 
the White filter light curve correlates with the flare at $T_0 + 98$~sec in 
the XRT with a lag of $\sim 5$~sec. The trace of the flare at $T_0 + 260$~sec 
is less evident. As mentioned in Sec. \ref{SECTION:uvot} 
there is a break in the $V$-band light curve (Fig. \ref{FIGURE:lightcurves}) 
with a lag of $\sim 5$~sec around this time, but it is less significant than 
the first flare.

As for the common gamma-ray and X-ray flares, Fig. \ref{FIGURE:lightcurves} 
shows that they are very close and correlate with each other with a lag of 
$\lesssim 1$~sec with $15-300$~keV BAT band. There is also a weak evidence 
of a peak close to $T_0 + 260$~sec and the correlation between this section 
of the BAT and XRT light curves gives a lag of $\sim 12 \pm 5$~sec.

\section{Broad-Band Spectral and Light Curves Variability}\label{SECTION:specvar} 
We fit the X-ray spectra using XSpec-12 \citep{A1996}. The model used was a 
power law spectrum with variable hydrogen column density in the host galaxy 
and a Galactic hydrogen column density fixed at 
$N_H^{MW} = 2.67 \times 10^{20}$~cm$^{-2}$. We adopted a redshift of 
$z = 3.082$ for this burst and its host galaxy. The photon counting data 
before the break - mainly the shallow slope regime - are well-fit by a single 
power-law with $\Gamma  = 1.59 \pm 0.06$ and an excess $N_H$ of 
$5.7^{+2.9}_{-2.7} \times 10^{21}$~cm$^{-2}$ at the redshift 
of the source. After the break the spectral slope is 
$\Gamma  = 1.73 \pm 0.08$. There is no evidence for a change in the column 
density across the break at $\sim 12200$~sec. The brightness of the X-ray 
afterglow permits the spectrum to be determined before, during, and after 
flares. From Fig. \ref{FIGURE:xrt_lc} it is clear that strong flares overlap 
the continuum X-ray emission. For this reason we have selected 4 intervals 
of $10$~sec during which the flares or the continuum are expected to be the 
dominant contributors to the X-ray lightcurve. The duration was selected 
such that there were enough events at later times - smaller fluxes - to obtain 
a statistically significant spectrum index and $N_H$. The first and the last 
intervals belong to the continuum in the region usually called tail emission. 
The last interval was positioned to be as far as possible from the transition 
region between tail emission and the shallow slope regime. The other two 
intervals are positioned near the peaks of the main flares. They allow to 
measure spectral differences between maximally different features of the 
light curve. The results are summarized in 
Table \ref{TABLE:spec_fits}. We discuss them later in this section along with 
the evolution of the X-ray hardness ratio.

\begin{table}
\caption{Spectral Fits to the $X$-Ray Data\label{TABLE:spec_fits}}
\begin{center}
\begin{tabular}{llll}
\hline \\
XRT-Flares & Time + $T_0$ (sec) & $\Gamma$ &
$N_\mathrm{H}$($10^{22}$cm$^{-2}$)\\
\hline \\
Before 1$^{st}$ flare  &  $79.5-89.5$   &  $2.09^{+0.26}_{-0.23}$  & $1.1^{+1.2}_{-0.9}$   \\
Peak 1$^{st}$ flare    & $91.5-101.5$   &  $1.34 \pm 0.13$         &    $< 0.92$    \\
Peak 2$^{nd}$ flare    & $257.5-267.5$  &   $1.60^{+0.13}_{-0.12}$   &  $1.1^{+1.0}_{-0.8}$  \\
After 2$^{nd}$ flare   & $394.5-404.5$  &   $2.00^{+0.42}_{-0.45}$   &  $< 0.76$ \\ 
\hline
\end{tabular}
\end{center}
\footnote*{$N_\mathrm{H}$ are at the redshift of the source $z = 3.082$. 
Galactic $N_\mathrm{H}$ is $2.6^\pm 0.05\times 10^{20}$~cm$^{-2}$.}
\end{table}

\begin{figure}
\includegraphics[height=6cm,angle=-90]{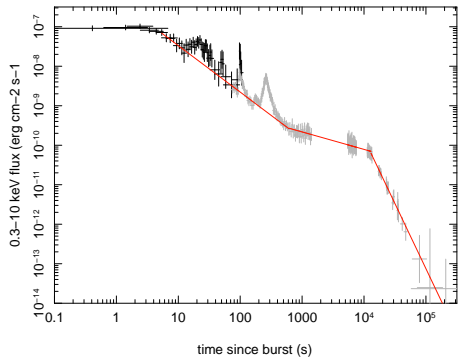}
\caption{The BAT light curve (black) extrapolated into the XRT band using the 
mean of the BAT and pre-flare WT spectra, and the XRT light curve (grey). The 
red curve shows the fit on the continuum component of the lightcurve from 
$T_0 + 5.41$ on. For $T_0 + 5.41$ to $T_0 + 635.96$ the slope is 
$\alpha = 1.17 \pm 0.04$, consistent at $1\sigma$ level with $\alpha_1$ in 
Sec.\ref{SECTION:xrt}. For the other segments the results are the same as 
what is reported in Sec.\ref{SECTION:xrt}.\label{FIGURE:xrtlc}}
\end{figure}

There are coincident BAT and XRT flares at approximately $100$~sec, so we 
performed a joint spectral fit to the BAT and XRT data during the flare. 
The best fit is 
obtained using the same model as described above for the XRT-only fits. It 
has $\Gamma  = 1.63 \pm 0.05$ and $N_H$ is consistent with the value that 
was determined using only the XRT data. Changing the model to a broken power 
law or a cut-off power law does not improve the fit. The joint BAT+XRT 
spectral slope is consistent with what was found for 
the XRT data, so we conclude that the X-ray flare is an extension of 
the flare seen in the gamma-rays. Fig. \ref{FIGURE:xrtlc} shows the 
BAT light curve extrapolated to the XRT band using the mean of the BAT and 
pre-flare WT spectra, along with the XRT light curve. In addition, it shows 
power-law fits on the different segments of the continuum component of the 
light curve. There is a smooth power-law decaying component, from the time 
at which the first 
gamma-ray peak begins to fall at $\sim T_0+5$~sec, until the end of the 
last major flare in X-ray at $\sim T_0 + 700$~sec. The decay slope of this 
segment is $\sim 1.17$. Flares, in both BAT and XRT, are superposed on this 
continuum. It is evident that flares observed by both instruments 
are in close relation, and therefore X-ray flares have the same origin as 
the prompt gamma-ray emission, presumably the internal shocks. As 
mentioned in Sec.\ref{SECTION:timevar}, the correlation of XRT and UVOT 
light curves shows an excess of optical emission during the flares. 
In conclusion, the late flares are simultaneously observed by all 
instrument on board of the \swift.

A confirmation of the hypothesis that flares have the same origin as the 
prompt gamma-ray can come from a direct comparison of their emission 
properties. There is a well known relation between the width of 
auto-correlation of peaks and the energy bands first discovered by 
Fenimore, \etal (1995) - the lower the energy band, the larger the width of 
the auto-correlation of a peak. We tested this relation for the BAT peaks 
and for 
the flare at $\sim T_0 + 98$~sec observed by both BAT and XRT. In all cases 
peaks have a wider auto-correlation for lower energy bands. Another suggested 
relation is the proportionality of the waiting time during a quiescent period 
and subsequent burst of radiation, evidence for a sort of accumulation 
of energy behaviour \citep{RM2001,RET2001,NP2002}. We can not confirm this 
relation for GRB 060607A. In fact, it seems that the amplitude of fainter 
peaks following the main peak at trigger time decreases according to a 
power-law, somehow shallower than continuum emission, regardless of 
the quiescent interval between them, see Fig. \ref{FIGURE:xrtlc}.

In order to estimate the shortest variability timescale of the prompt and 
flares we use their Fourier transform and auto-correlation. The shortest 
variation timescale distinguishable from a white noise for the prompt 
gamma-ray from $T_0 - 5$~sec to $T_0 + 95$~sec is $\sim 2.5 \pm 1$~sec in 
all bands. For the flare (peak) at $\sim T_0 + 98$~sec, the minimum 
variability distinguishable from noise is close to the main peaks, 
$\sim 3 \pm 1$~sec, using the total light curve for the BAT in $15-300$~keV 
and for the XRT data in $0.3-10$~keV. The minimum variability of the flare 
at $\sim T_0 + 260$~sec is $\sim 7 \pm 5$~sec, longer than previous peaks. 
Variation in the emission can be due to the inhomogeneity in the shells 
and/or their limited size. In this case one expects that with the expansion 
of the fireball or what rests from it, the density variation dilutes. 
Therefore, what we expect is an increasing minimum time variability with 
time, which is exactly what we are observing. However, as with expansion 
the signal becomes fainter and the variations become more difficult to 
observe, the increase in minimum variability scales is a consequence of 
both effects. 

As the speed of the passage of the shock front through a shell is limited 
to the speed of light, these timescales can be translated to the distance 
traveled by a relativistic ejecta. Therefore, they 
constrain the initial size and/or variability scale of the fireball, and 
the distance to the central engine to $\gtrsim 3 \times 10^{10}$~cm in the 
rest frame of the engine. This is consistent with the estimation of the 
model explained in the Appendix \ref{appendixa} and 
applied to the data in Sec.\ref{SECTION:modelling}. The timescales are 
also another confirmation of the same origin for the prompt emission and 
flares seen in X-ray and optical bands. The increasing timescale of 
variations is consistent with the expansion of what rests from the prompt 
shell (fireball) after internal shocks and its coalescence with other shells.


As for the continuum emission, there is no evidence of a steep initial 
decline as would be expected from a high-latitude emission. One 
possibility is that the external shock phase with ISM/surrounding material 
began very early and smeared the high latitude emission. Another possibility 
is that the continuous emission and the preceding peaks in the prompt 
emission have the same origin; with the continuous emission due to the 
decaying tail of the emission from the prompt shell after its main collision.
As explained earlier in this section, a power-law fits the extrapolated BAT 
light curve to the XRT energy band. The smoothness of the joint lightcurve 
is a likely evidence that the initial smooth, sometimes very steep decay 
observed in the X-ray light curve of many GRBs is directly related to the 
prompt emission and hence the term {\it tail emission} is a correct 
expression for this regime. As for the physical processes 
involved, it is possible that the magnetic field in the 
coalesced shells has a relatively long lifetime and electrons are accelerated 
and support a synchrotron emission at a lower rate well after the end of the 
collision between shells. See also Sec. \ref{SECTION:modelling} for more 
details. There is however a caveat in this argument. The extrapolation of 
the BAT light curve to the XRT uses the average spectrum slope observed by 
these instruments in their corresponding bands. Therefore, the smoothness of 
the joint light curve may be due to the way it is 
calculated, thus the argument about common origin of the prompt and tail 
emission becomes doubtful. Nonetheless, we show in Appendix \ref{appendixb} 
that if the time evolution of the BAT spectrum index is taken into account, 
it is highly improbable to obtain such a smooth common light curve due to 
averaging or by chance, and one should observe a deviation between 
extrapolated BAT light curve and what is observed by the XRT if there is not 
an intrinsic relation between the prompt and the tail emissions.

\begin{figure}
\includegraphics[height=6cm]{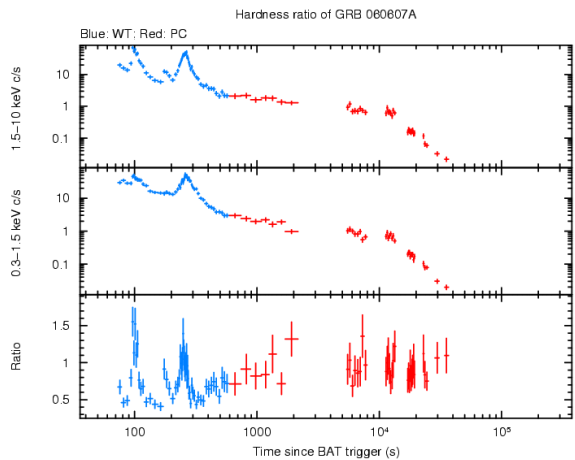} \caption[xrthrratio.eps]{The 
upper and middle panels show the count rate measured by the XRT in 
$1.5-10$~keV and $0.3-1.5$~keV respectively. The lower panel shows the 
evolution of the hardness ratio, the ratio of the flux measured in the $1.5-10$~keV band to that measured in the $0.3-1.5$~keV band. Darker 
points are observations in WT (Window Timing) mode and lighters in PC 
(Pulse Counting) mode. \label{FIGURE:lc_hr}}
\end{figure}

Fig. \ref{FIGURE:lc_hr} shows the evolution of the X-ray hardness ratio: 
$HR_X \equiv C(1.5-10$~keV$)/C(0.3-1.5$~keV$)$, where $C$ is the count rate 
uncorrected for the absorption. Both of the X-ray flares are significantly 
harder than the underlying decay, consistent with the spectral indices 
reported in Table \ref{TABLE:spec_fits}. Moreover, comparing the spectral 
index of the spectrum of the first and the last time intervals reported in 
this table, and their hardness ration shown in 
Fig. \ref{FIGURE:lc_hr}, it seems there is no evidence for large spectral 
evolution in the continuum emission until entering to the shallow slope 
regime. The hardness ratio during each flare tracks the luminosity; that is, 
the 
hardness increases as the flare brightens, and decreases as the flare fades. 
Another interesting observation from this plot is a gradual rise of the 
hardness ratio from the beginning of the shallow regime to a roughly 
constant plateau that does not decline even after the sharp break at 
$\sim 1.2 \times 10^4$~sec. If the break was achromatic such a behaviour 
was expected, however the break is not achromatic. It can be due to 
a decrease in the number of emitters - accelerated electrons - 
and hardening of their spectrum, and/or a stronger magnetic field. In 
Sec. \ref{SECTION:modelling} we discuss these issues more in details.

The fluence in gamma rays (15-150 keV) during the first pulse 
($T_0 - 24$ to $+12$ sec) is $(1.38 \pm 0.05) \times 10^{-6}$~erg~cm$^{-2}$ 
while the corresponding fluence between $24$~sec and $102$~sec is 
$(1.22 \pm 0.05) \times 10^{-6}$~erg~cm$^{-2}$. There is no flux in the 
$15-150$~keV band after $102$~sec. Therefore, unlike the first flare the 
second XRT flare has no (or a marginal) BAT counterpart. This suggests that 
the two XRT flares are produced by distinct shells colliding with the remnant 
of the prompt shell. 

It seems that both main flares were preceded by fainter but harder flares, see 
Fig.\ref{FIGURE:xrtlc}. This repetition can be explained if two late shells 
successively pass through the remnant of the prompt shell. In this case the 
observed emission is the tomography of the prompt. The small and large 
successive flares can be the result of collision between the late shell, 
slower shocked material, and a leftover unshocked material from the prompt 
shock with a larger Lorentz factor that moves ahead of the first component. 
A similar configuration of flares is also visible in the X-ray light curve 
of GRB 070107\citep{STAM2007}. Assuming that the late shells have roughly the 
same dynamical properties - Lorentz factor, density, etc. - the softness of 
the second flare could be due to the expansion and slowdown of the unshocked 
remnant. The expansion of the late shell can also be in part responsible 
for the softer radiation during collision, which in this interpretation 
had necessarily happened at larger distances from the central engine.

The observed X-ray decay slope after flares and up to $\sim T_0 + 600$~sec 
when the shallow regime began, is $\alpha_X = 1.09 \pm 0.04$. The spectral 
index of the X-rays in the same interval is 
$\beta_X = \Gamma - 1 = 0.64 \pm 0.07$. If the emission 
in this interval is due to an external shock with ISM/circumburst material, 
the closure relationships can be used to determine the density distribution. 
We consider two cases: a constant density circumburst medium and a 
wind-stratified circumburst medium. The closure values are given in 
Table \ref{TABLE:closure}. The case that gives a closure closest to zero 
is one with a constant density medium and the cooling frequency above the 
X-ray band at $600$~sec. Further evidence for the cooling break being 
above the X-ray band in this time interval is that the optical decay has a  
slope of $\alpha_{opt} = 0.9 \pm 0.06$, close to the X-ray slope at 
$1\sigma$ and consistent with it at $2\sigma$ level. 
If this scenario is correct, the electron index is predicted to be 
$q = p + 1 \sim 3.5$, consistent with what is seen in many other GRBs 
\citep{SHE2006}. The optical decay does not change between $\sim 600$~sec and 
$\sim 20000$~sec after the BAT trigger, so the cooling break must be above 
optical frequencies during this period.

As for what we can learn about the gas and dust content of the host galaxy 
of GRB 060607A, the fitted neutral hydrogen column density in the host 
along the line of sight to the burst is consistent with 
$N_H = 1.6 \times 10^{22}$~cm$^{-2}$, which implies a high extinction in 
the host galaxy. \citep{PS1995} find $N_H = (1.79 \times 10^{21}) A_V$ for 
the conversion between hydrogen column density and extinction in the Milky 
Way. If this relationship holds for the host galaxy, then $A_V = 8.9$~mag 
in the host. While the optical data does not rule out such a high extinction 
there is no evidence for it either. If we assume an $N_H/A_V$ ratio like 
that in the SMC of $N_H = (15.4 \times 10^{21}) A_V$ (from equation (2) and 
Table 2 of \citep{P1992} then $A_V$ in the host is $1.0$~mag.

Gas-to-dust ratios similar to that of the SMC have been observed for several 
GRB host galaxies, such as GRB 000301C \citep{JFG2001}, 
GRB 000926 \citep{FGD2001}, GRB 020124 \citep{HMG2003}, and 
XRF 050416A \citep{HBG2007}, see also \citep{SFD1998,KKZ2006}. Detailed 
investigation of 7 \swift bursts \citep{S2007} shows that only in one case 
the extinction is best modelled by Milky Way gas-to-dust 
ratio, and for other cases the extinction is more similar to SMC. Giving 
the fact that GRB060607A was observed in all the rest frame bands redder 
than Ly$\alpha$ suggests that it may be reasonable to assume that the host 
galaxy of GRB 060607A has a high gas-to-dust ratio. This could indicate that 
star formation in the host is fairly recent and there has not been enough 
time for large amounts of gas to be processed into dust. Alternately, a high 
ratio could be indicative of dust destruction in the vicinity of the 
progenitor by the burst itself \citep{WD2000,PKG2003,W2007}.

%
%
%
%
%
%
%
%
%
%
%


There is no evidence for a jet break in the optical light curve out to 
$\sim 20000$ sec after the BAT trigger. This, and the isotropic energy of 
the burst, can be used to put a lower limit on the opening angle of the jet, 
and the total gamma-ray energy of the burst \citep{R1999,SPH1999,FKS2001}. 
Assuming a single jet, the lower limit of the jet opening angle for 
GRB 060707A is:

\bea
\theta_j & \ge & 0.161 \biggl (\frac{t_j}{z+1}\biggr )^{3/8} 
\biggl (\frac {n \eta_\gamma}{E_{iso}} \biggr)^{1/8} = \nonumber \\
&& 0.025 \left( {\eta_\gamma \over 0.2} \right)^{1/8} \left( 
{n \over 0.1} \right)^{1/8} rad,
\eea

where $\eta_\gamma$ is the efficiency of converting energy in the ejecta into 
gamma rays, and $n$ is the particle density in cm$^{-3}$. The corresponding 
energy in gamma rays, using $E_{iso}$ obtained in Sec. \ref{SECTION:bat} in 
the rest frame of the engine and corrected for the beaming is 
$E_{\gamma} \geqslant 3.3 \times 10^{49}$~erg.

\begin{table}
\caption{This Table lists the closure relationships for various
assumptions about the location of the cooling frequency and the nature
of the circumburst medium.  A closure value of zero indicates
agreement with the predictions of each case.\label{TABLE:closure}}
\begin{center}
\begin{tabular}{cclc}
\hline \\
Model & Environment & Closure & Value \\
\hline \\
$\nu_X < \nu_c$ & ISM  & $\alpha - 3/2\beta$       & $-0.13$ \\
                & Wind & $\alpha - 3/2\beta + 1/2$ & $+0.63$ \\
$\nu_c < \nu_X$ & ISM  & $\alpha - 3/2\beta - 1/2$ & $-0.37$ \\
                & Wind & $\alpha - 3/2\beta - 1/2$ & $-0.37$ \\
\hline
\end{tabular}
\end{center}
\end{table}

\section{Modelling}\label{SECTION:modelling}
In this section we apply a reformulation of the internal/external shock model 
for the prompt and afterglow to the data in order to estimate some of the 
parameters of the burst. This formulation is based on a simplified 
ultra-relativistic radiative shock model with one synchrotron emitting 
shocked layer but more detailed parametrization of the physical processes 
and their time variation (Ziaeepour, in preparation). A summary of the model 
and its main results are given in the Appendix \ref{appendixa}. 

\subsection{Methodology}\label{SECTION:method}
A major difficulty in understanding the behaviour of GRBs is that the main 
ingredients of the shock i.e. the electric and magnetic fields, and the 
distribution of electrons varies in a complex manner with time during the 
evolution of the microphysics and the dynamics of the shock 
\citep{WA2004,BO1996,WD2000,RKD2006,RBD2006}. The commonly used power-law 
parametrization 
with constant coefficients and indexes is not able to explain the complex 
behaviour of quantities and thereby the synchrotron emission. Adding the 
microphysics of the shock to the formulation of the shock dynamic is also 
too difficult as in most cases there is no analytical expression for their 
evolution, or for the evolution of the electric and magnetic fields  or 
distribution of electrons. As we will 
explain in more details, we found that the best way to estimate 
parameters and to explain the behaviour of the light curves is to divide them 
into sub-regimes that can be explained separately by a simple power-law 
parametrization. On the other hand, this simplified model still requires 12 
parameters and 
thus fitting such a complex model to the data is not trivial, even after 
sub-dividing it into separate regimes. The degeneracy between parameters 
and the 
instability of numerical fitting can lead to confusing results. Moreover, 
at the current level of our knowledge about GRBs and the precision of 
available data, even a rough estimation of parameters permit a better 
understanding of the nature of these elusive objects and can be considered 
as an achievement. 
Therefore, rather than fitting the data, we determine the predictions of the 
model through numerical simulations with prefixed parameters. Then, 
we estimate the set(s) of parameters that best reproduce the 
behaviour of the data.

\begin{figure}
\begin{center}
\includegraphics[width=7cm]{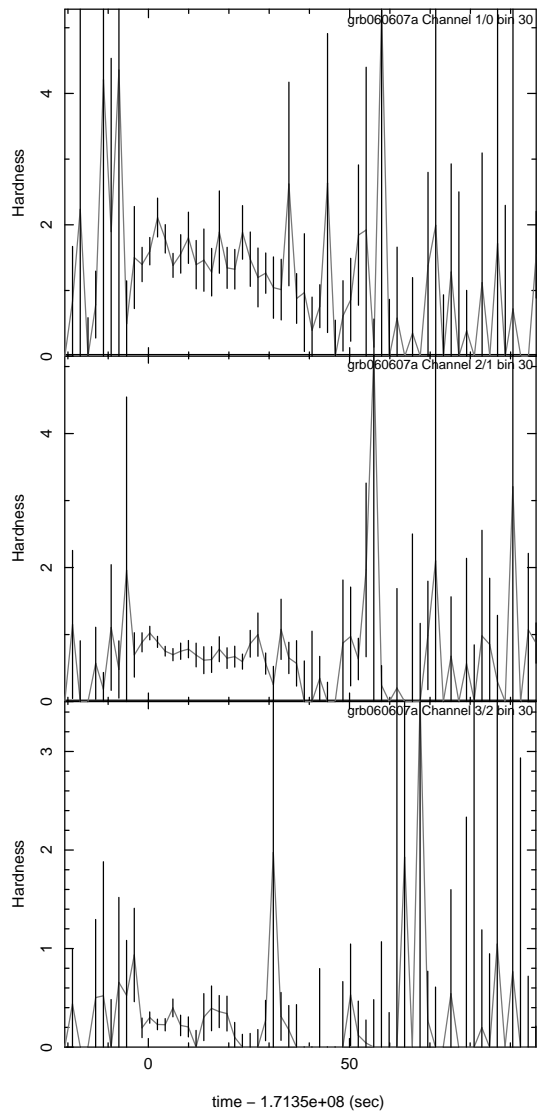}
\caption{Hardness ratios of the main peak. From top to bottom: $HR_{10} \equiv
C(25-50~keV) / C(15-25~keV)$, $HR_{21} \equiv C(50-100~keV) / C(25-50~keV)$, 
and $HR_{32} \equiv C(100-300~keV)/ C(50-100~keV)$. For calculation of the 
ratios data from the $64$~msec binned light curve are used and rebinned by a 
factor of 30. \label{FIGURE:bathardness}}
\end{center}
\end{figure}

\subsection{Modelling and interpretation of the BAT data}\label{SECTION:batmodel}
To interpret the light curves of the GRB 060607A, we begin with the hardness 
ratios of the BAT bands. The hardness ratios of the first main peak are 
shown in Figs. \ref{FIGURE:bathardness}. The initial rapid rise is 
consistent with an exponential growing to a saturated value of $\epsilon_e$ 
and $\epsilon_B$, respectively, of the fraction of kinetic energy 
transferred to accelerated electrons and to magnetic field. This is also in 
accordance with simulations of the formation of magnetic 
field \citep{Y1994,WA2004}, parallel instabilities \citep{RKD2006} in the 
relativistic shocks, and acceleration of particles in a Fermi process 
\citep{BO1996,RBD2006}\footnote{For simplicity in all the simulations 
discussed here we use a power-law with constant index distribution for 
electrons.}. After this transient time, fields settle into a quasi-stationary 
regime in which they decline according to a power-law. Comparison with 
simulations show that for this burst 
$\eta \equiv 2 \alpha_e + \alpha_B/2 \sim 3.5 \pm 0.5$ for a power-law 
electron number distribution index $q \sim 3.5 \pm 0.5$, where $\alpha_e$ 
and $\alpha_B$ are time/radius dependence of $\epsilon_e$ and $\epsilon_B$, 
respectively. The value of $q \sim 3.5$ (or equivalently $p \sim 2.5$) is 
close to $p \sim 2.2-2.3$, the {\it universal} value suggested by 
simulations \citep{KI2000,ACH2001}. On the other hand, the distribution 
of $p$ obtained from the observations of GRBs and other 
relativistic sources does not show a universal average and has a relatively 
large deviation around the mean value \citep{SHE2006}. This can be the 
evidence for more complex outflow behaviour such as: a complex geometry of 
the magnetic field and shock, plasma 
currents and turbulence, non-Fermi acceleration mechanisms, or dominance 
of the Poynting flux. In conclusion, although the $p$ value obtained here 
is typical, it can not give more information about the underlaying processes.
The estimated value of characteristic synchrotron energy of the 
electrons with minimum Lorentz factor $\Gamma$ is 
${\mathcal E}_m \equiv h \nu_m \sim 12 (1 + z)$~keV $\sim 50$~keV in the 
rest frame of the engine. 

These estimations are based on the hardness 
ratios $HR_{10}$ and $HR_{21}$. The highest energy band hardness 
$HR_{32} \equiv C(100 - 300 keV )/C(50 - 100 keV )$ is much noisier 
than $HR_{10}$ and $HR_{21}$. However, it seems that the electron spectrum 
should be much steeper to explain the observed low value of 
$HR_{32} \sim 0.4$ in place of the expected value of $\sim 0.65$ from 
simulations for $q$ and $\eta$ as mentioned above. We interpret this as an 
upper limit cutoff or steepening in the 
electron distribution. Simulations of electron acceleration by 
shocks \citep {ah2006} also confirm a steeper distribution at high 
energies. A change in the spectrum of electrons has been also invoked as 
an explanation for the chromatic break in the X-ray and optical 
afterglows \citep{wl2002,M2007}. The observation of such a break/steepening 
in electron energy distribution during the prompt emission, when the shock 
is much stronger, shows that the popular assumption of a power-law 
distribution is too simplistic and far from reality. The estimated range of 
parameters includes both uncertainties and degeneracy 
between them. The initial (maximum) value of the hardness ratios depends 
mainly on $q$, their decreasing rate depends on $\eta$ and coefficient 
$F$ defined in \ref {EQUATION:defeps}, and there is a significant degeneracy 
between these quantities.

From the slope of the hardness ratios in the quasi-steady regime and the 
lag between the BAT energy bands, we can estimate the coefficient $F$ and 
the distance to the central engine. The lags between the BAT bands 
are presented in Table \ref{TABLE:lag}. Considering the peak time after 
the initial rise as the beginning of the power-law decline regime for 
$\epsilon_e$ and $\epsilon_B$, and by comparing lags with simulations, we 
conclude that $F \sim 12 - 18$, $\Delta \epsilon_{02} \sim 6 \times 10^{-6}$, 
and $\Delta \epsilon_{12} \sim 3 \times 10^{-6}$, where 
$\Delta \epsilon_{ij}$ is the difference between $\epsilon$ defined in 
Eq. \ref{EQUATION:defeps} at peak time for bands i and j. Including 
uncertainties both in data and comparison with simulations, we estimate 
the initial distance to the central engine as $r_0 \sim 10^{12}$~cm. This is 
roughly in the middle of the distance range predicted for the internal shocks 
\citep{RM1994}\footnote {In all the calculations here we assume that the 
bulk Lorentz factor of the fireball $\gamma_{bulk} \gg 1$ and 
$\beta_{bulk} \sim 1$.}.

\begin{table}
\caption {BAT lags\label{TABLE:lag}}
\begin{center}
\begin{tabular}{cr}
\hline \\
Band & Lag (ms), entire burst \\
\hline \\
     0-2 &  $621.90 \pm 57.0$ \\
     1-2 &  $252.18 \pm 25.7$ \\
     3-2 & $-556.73 \pm 57.0$ \\
\hline
\end{tabular}
\end{center}
\end{table}

Knowing $F$ and $r_0$, we can also estimate the relative Lorentz factor of the 
shells. In the radiative shock model assumed here colliding shells coalesce. 
Therefore, at the end of the shock, $\gamma \rightarrow 1$. 
Using Eq. \ref{EQUATION:defeps}, $\gamma_0^2 \approx 1 + F \epsilon_f$ where 
$\epsilon_f$ corresponds to the value of $\epsilon$ at the end of the 
coalescence. It is not very evident what time should be 
used to determine $\epsilon_f$, because substructures/overlapping peaks 
can be due to separate shells or from density inhomogeneities in the same 
shell. Moreover, the limited sensitivity of the BAT can smear the real 
collision time. Nonetheless, if we assume that the detected duration of the 
peaks corresponds to the main part of the collision, an estimation of 
$\gamma_0 \sim 1.5 - 3.5$ for overlapping peaks from $\sim T_0 - 5$~sec 
to $\sim T_0 + 40$~sec can be made. This includes uncertainties in the 
parameters and the duration of the collision. The bulk Lorentz factor of 
the ejecta however can not be determined from the prompt without knowing 
all the parameters such as the density of shells, magnetic fields, etc. The 
reason is the fact that the physics of shock depends only on the relative 
Lorentz factor, and the detected radiation is just boosted by the bulk 
Lorentz factor to the observed energies. Other BAT peaks are too weak to 
permit a detailed analysis of their corresponding shock.

\subsection{Modelling and interpretation of the XRT data}\label{SECTION:xrtmodel}
If we neglect flares that are superimposed on the early XRT light curve 
shown in Fig. \ref{FIGURE:xrt_lc}, the power-law component from 
$\sim T_0 + 73$~sec to $\sim T_0 + 700$~sec has a slope of 
$\alpha_1 \sim 1.09$ and spectrum index of $\beta_1 \sim 0.64$. These indices 
do not 
satisfy the relation $\alpha_1 = \beta_1 + 2$ for high latitude emission. 
Therefore, we interpret this section of the X-ray light curve as the tail 
emission 
from energy dissipation in the prompt shell after the termination of the 
coalescence. Investigation of the hardness ratio of the X-ray 
bands $C(1.5-10~keV )/C(0.3-1.5~keV )$ shows that at this time the ejecta has 
significantly cooled, consistent with the results of Sec. 
\ref{SECTION:specvar}. In fact, we can go further and determine 
${\mathcal E}_m$ at the beginning of the observations after {\sl Swift\/} 
slewed to this burst.

Assuming synchrotron radiation as the source of the observed X-ray, 
according to our model, ${\mathcal E}_m$ evolves 
as\footnote {Here we neglect the change in the bulk Lorentz factor. This 
happens mainly during the coalescence.}:

\begin{equation}
{\mathcal E}_m(r) = {\mathcal E}_m(r_0)
                    \biggl( \frac {r}{r_0} \biggr)^{-\eta}
                    \biggl( \frac {\gamma^2}{\gamma_0^2} \biggr)^{\frac{5}{4}}
\label{EQUATION:emevol}
\end{equation}

After the coalescence of the two shells $\gamma = 1$, and therefore at 
$\sim T_0 + 73 / (z+1)$~sec in the source frame when the coalescence of the 
shells has been already 
finished, the second term in Eq. \ref{EQUATION:emevol} is fixed to 
$(1/\gamma_0^2)^{5/4}$. Using $r_0$, ${\mathcal E}_m(r_0)$, and 
$\gamma_0 \sim 2.5$ - the mean value in the range derived from BAT data - we 
find that ${\mathcal E}_m$ has been reduced by a factor of $\sim 0.026$ to 
${\mathcal E}_m \sim 0.3(1+z)~keV \sim 1.2~keV$ . Our simulations show that 
the observed X-ray hardness ratio (Fig. \ref{FIGURE:lc_hr}) of 
$\sim 0.5 \pm 0.2$ just before the first flare is consistent with the 
reduced value of ${\mathcal E}_m$ mentioned above, only if the spectrum 
of the electrons also has steepened from $q \sim 3.5 \pm 0.5$ to $q \sim 5$. 
The steepness of this slope may signify the failure of the simplified model 
without an upper limit or high energy break in the energy distribution of 
accelerated electrons. As mentioned above, this assumption is not realistic 
and can compromise the interpretation of the data.

The value we obtain for the decay index of the fields $\eta \sim 4$ is 
also slightly higher than its value after the steady state regime during 
the main peak in the prompt gamma-ray emission $\eta \sim 3.5$. Both the 
increase in $\eta$ and electron distribution index $q$ are consistent 
with the cooling of the shell in this regime. What is remarkable is that 
we do not see a faster decay of the fields after the collision and coalescence 
of the shells both terminated. This means that the coherent bulk and field 
structures formed during the collision have relatively long lifetimes and 
do not disappear immediately after the end of collision.


Fig. \ref{FIGURE:lc_hr} shows the hardness ratio of the X-ray bands for 
multiple flares detected by the XRT. The flare at $\sim T_0 + 98$~sec was also 
observed by the BAT, and there is likely a small but delayed trace of it in 
the UVOT light curve, as explained in Sec. \ref{SECTION:timevar}. 
A noticeable difference between the flare hardness ratio in the X-ray and 
the gamma-ray emission is that the hardness ratio of the X-ray flares follows, 
sometimes with lag or lead, the X-ray light curve, whereas, the hardness 
ratio of the gamma-ray emission, rises to a plateau and then declines slowly 
until the end of the spike\footnote{Note that we compare the behaviour of 
the X-ray hardness ratio of the main BAT peak with XRT flares. Although 
the first XRT flare of GRB 060607a was also observed by BAT, it is too weak 
to permit the calculation of a meaningful hardness. Nonetheless, the 
similarity of the profiles in Fig. \ref{FIGURE:bathardness} suggests that 
the XRT hardness which is calculated for energy bands just less than one 
order of magnitude lower should most probably be a continuation of the same 
type of behaviour.}. A strong correlation is usually observed in flares 
where flares that brighten also harden \citep{GOA2007}. This is 
interpreted as the consequence of the spectrum evolution and a decreasing 
break energy. However, in the case of GRB 060607a, a high energy break  
was not observed in the XRT or BAT spectrum. Even without a break, 
the spectral evolution during a flare could still produce this correlation. 
Nonetheless, rapid variation of the hardness ratio of flares in GRB 060607a 
does not seem to be a typical behaviour of all bursts, 
see for instance the case for GRB 061121 \citep{P2007}. One explanation could 
be a strong absorption of the soft band photons which flatten the soft X-ray 
light curve. However, the flux hardness ratio curve in which corrections 
are made 
for the absorption has a very similar behaviour, and therefore absorption can 
not be the cause. Another possibility is that the magnetic field and 
electron acceleration during the collision of the late shells with the 
remnant of the prompt shell did not achieve to rise to the steady state 
regime and decayed exponentially once the late shell either coalesced with 
the prompt or passed through it. Our simulations also confirm this possibility.

As for the physical reasons for such behaviour, one important factor can be 
the heating of the prompt shell after its collision. It is well known 
\citep{WA2004,WD2000} that the development of a coherent magnetic and 
electron acceleration is weaker when the shock medium is hot and particles 
have a significant momentum in the direction perpendicular to the direction 
of the boost/bulk movement. As the hardness ratio in the flares does not 
seem to arrive to a steady state regime, and we do not know the size and 
other characteristics of the late shells, we can not estimate shock 
parameters for them as we did for the prompt gamma-ray peak. Nonetheless, 
from our simulations using $q$ and $\eta$ similar to 
the prompt and continuous component of the X-ray light curve, we conclude 
that the shock is soft, consistent with weak or non-observation in the 
BAT bands.

Like other bursts the most difficult part of the X-ray light curve to 
explain is the shallow slope regime. Various processes are suggested, such 
as: refreshed shocks by late shells \citep{RM1998,SM2000}, continuous energy 
injection to the fireball (prompt shell) \citep{DL1998,RM2000,Y1994}, and 
variation of the microphysics of what remains from the prompt shell 
\citep{ITY2006}. The first two processes seem to have an energy problem; 
they need much larger efficiency for gamma-ray emission than 
expected \citep{ITY2006}. In addition, the uniform slope of this regime 
in all bursts does not look like a phenomenon relaying on the random 
injection of late/slow shells. The abrupt break at the end of this 
regime - especially in the case of GRB 060607A - needs a sudden stop of 
both energy injection and radiation that seems unphysical. By contrast, 
a change in the microphysics and energy dissipation of the prompt remnant 
and later shells seems a more reasonable cause. In fact, we see such a 
late time slow evolution of the intensities in the simulations. However, 
we do not see an abrupt break. Assuming that the late break in the X-ray 
light curve is not the jet break as it is not achromatic, its absence in 
simulations may be due to the limited precision of the numerical 
simulations and the simplicity of the model. Moreover, the 
fact that our analytical approximations are valid only for $\epsilon < 1$, 
and therefore, simulations are limited to this linear regime can be another 
factor influencing the lack of a break in the simulations. If this 
interpretation is correct, parameters of the shell are similar to the tail 
emission with a steeper slope. The X-ray hardness ratio is however higher 
than the tail emission see Fig. \ref{FIGURE:lc_hr}. Energy injection by 
flares to the prompt shell can be the reason for the hardening of the 
synchrotron emission. Another explanation can be the onset of a forward 
external shock by the ISM or circumburst material at a distance 
$\sim 10^{14}$~cm from the central engine.

The question which arises here is: At these late times, what keeps the 
coherent magnetic field and the electron acceleration in the shell going? 
Do we need a continuous fall of circumburst material or arrival of late 
shells (i.e. energy injection) to the shell to keep the radiation level 
high? Are the stock of electrons and residual magnetic field enough to 
keep the low flux of the late time radiation? In the first case either 
the circumburst material should exist at all distances after the prompt 
collision, or there must be a discontinuity in the emission between 
the end of the internal shock and the onset of the external forward shock. 
The same type of argument is relevant for the continuous energy injection. 
No discontinuity has been observed by {\sl Swift\/}. Only a detailed 
knowledge of the origin of the ejecta, the shock, the state of the matter 
in the remnant after the prompt shock termination, and the surrounding 
material can clarify these issues. 

\subsection{Modelling and interpretation of the UVOT data}\label{SECTION:uvotmodel}
Finally we try to explain the optical light curve of GRB 060607A which has 
interesting behaviour rarely seen in other bursts. In general, optical light 
curves of {\sl Swift\/} bursts within the first few hundred seconds after 
the trigger, have a broad range of temporal indices, mostly consistent with 
decaying behaviour. After approximately 500s, all light curves are decaying 
(Oates et al. in prep). In few bursts including GRB 060607A (see 
Figs. \ref{FIGURE:lightcurves} and \ref{FIGURE:uvot_flare}), 
GRB 060418 \citep{MVM2006}, and GRB 070616 \citep{ST2007} an initial rise 
in the optical light curves is 
detected in all filter from blue to infrared. A number of interpretations 
for this behaviour have been put forward: The onset of external 
shock\citep{NH2007}, forward shock with ISM \citep{MVM2006}, deceleration 
of the forward shock\citep{S1997}(Oates et al. in prep). Here we argue that 
the 
most plausible reason for this behaviour is the gradual cooling and energy 
dissipation in the prompt shell and the entrance of ${\mathcal E}_m$ to 
optical bands. The relatively shallow slope of the early X-ray light curve, 
$\alpha = 1.09$ may be the consequence of a slow evolution of ${\mathcal E}_m$ 
which permitted the detection of its transition through the optical bands.

\begin{figure}
\includegraphics[height=6cm,angle=-90]{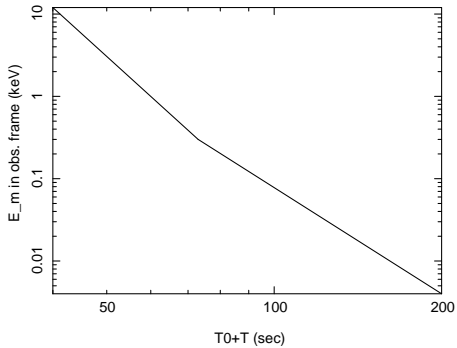} \caption{Estimation of the 
time evolution of ${\mathcal E}_m$ in observer frame, from the end of the 
gamma-ray peak until the optical peak according to the model explained in the 
Appendix \ref{appendixa}.\label{FIGURE:em}}
\end{figure}

Using Eq. \ref{EQUATION:emevol} and the argument about the evolution of 
$\gamma^2/\gamma_0^2$, we can estimate ${\mathcal E}_m$ at the peak time in 
optical light curve $t_{peak}$ $\sim T_0 + 200/(z+1)$~sec in the source 
rest frame, and we obtain 
${\mathcal E}_m \sim 4(1 + z)~eV = 16~eV$. This corresponds to $\sim 440$~nm, 
with in the optical blue band for the observer. Therefore, we interpret the 
maximum of the 
light curve as the time where the characteristic synchrotron emission for 
the least energetic electrons enters to the optical bands. The approximate 
time evolution of ${\mathcal E}_m$ according to the model used here is 
summarized in Fig.\ref{FIGURE:em}. 

The rising slope of the optical light curve is 
$\alpha^{rise}_{opt} \sim 2.3 \pm 0.3$. According to the model, the slope in 
this regime is $q/2 - 1/6$. With the estimated $q \sim 5$ from the X-ray data, 
the observed value is consistent with the interpretation of the peak as the 
passage of ${\mathcal E}_m$ to optical bands. Therefore, another component - 
external forward or reverse shock - is not required to explain the 
observations. Although there are UV observations for this burst, they do not 
begin until hundreds of seconds after the rise. Therefore it is not possible 
to see if they peak earlier than longer wavelengths, as would be expected 
from this interpretation. On the other hand, 
observations by the PROMPT telescope \citep{NH2007} in $B$, $g'$, $r'$, and 
$i'$ are consistent with this interpretation. In fact the energy difference 
between $B$ and $i'$ bands with respect to the energy difference between 
X-ray and optical is very small, and therefore the expected time shift between 
their arrival to maximum is very small and difficult to detect with the 
time resolution of ground telescopes. Only UVOT in event mode could be able 
to directly detect the shift, but it can not simultaneously observe with 
multiple filters. It may be possible to detect a small shift in the maximum 
by fitting the peak, but \citep{NH2007} did not attempt this. Nonetheless, 
in Fig. 1 of \citep{NH2007} which, includes the $H$ band data from the REM 
observations \citep{MVM2006}, shows a small delay between the peak time in 
the $H$ band at $180 \pm 6$~sec \citep{MVM2006} with respect to higher 
energy bands. We should also mention that 
flares observed in gamma- and X-ray bands are marginally distinguishable 
in the UVOT observations. This shows that as expected, the synchrotron 
emission from the flares is much harder than the cooling tail emission, and 
consequently no significant optical excess is observed during the 
X-ray flares. The lack of evidence for a significant external shock 
in the early low energy bands emission makes the claim of determination of 
the bulk Lorentz factor \citep{MVM2006} unlikely, as it is based on the 
assumption that this peak is the result of a shock on the ISM. In 
this case $\gamma_{bulk}$ is the same as collision $\gamma$ because ISM is 
roughly at rest with respect to the source. But according to our results the 
peak is related to the prompt shock and so the ISM/circumburst material had no 
significant contribution in its formation and properties.

The rest of the optical light curve follows a power-law decay at least until 
$\sim T_0 + 25000$~sec without any change in the slope. The magnitude 
limits at later times are just brighter than the expected fluxes from a 
power-law decay and we can not claim that any late break in optical/UV bands 
has been observed. The flattening or slight brightening from 
$\sim T_0 + 1000$~sec to $T_0 + 2000$~sec, (see Fig. \ref{FIGURE:uvot_data}) 
is probably related to the weak X-ray flare that occurred during this period.

Although the number of data points in the light curve in this regime is 
very limited, one can speculate that the slope of the decline is slightly 
steeper in longer wavelengths. This can be interpreted as the hardening of the 
emission, probably due to fireball encountering the ISM/circumburst material. 
In other 
words, it seems that a weak external shock starts a few thousands of seconds 
after the trigger and refreshes the prompt shell. Regarding just optical 
data, this claim seems very speculative. However, the hardness ratio of 
X-ray bands is definitely increasing (Fig.\ref{FIGURE:lc_hr}). This 
increases the hardness ratio to $\sim 1$, i.e. ${\mathcal E}_m$ moves to 
the X-ray band but with a smaller density for high energy electrons. As 
the shell is already decelerated, a few thousands of seconds later at 
$\sim T_0 + 1.2 \times 10^4$~sec the additional mass of the coalesced 
material from ISM breaks the coherence of the magnetic field and the 
synchrotron emission at higher energy bands begins a steep power-law decay 
due to a break in the spectrum of the electrons. This does 
not affect the emission in the optical bands from slower electrons. In this 
sense one can say that the X-ray and optical emissions come from separate 
components. It is also possible that at least part of the optical emission 
comes from the heated-shocked material without the presence of a coherent 
magnetic field. In conclusion, the break is not likely to be 
associated with the jet break, i.e. when $1/\gamma_{bulk} > \theta_{jet}$, 
and the relativistic collimation is removed.

\subsection{Alternative interpretations}\label{SECTION:altmodel}
To end the discussion, we consider some of the alternative interpretations 
of the optical light curves. First we consider the possibility that the 
optical emission is produced by a forward or reverse external shock, and 
thus has a different origin than the prompt gamma ray and the early X-ray 
emission. Considering first the forward shock case, the peak of the optical 
emission marks the onset of the afterglow where the peak time of the 
optical emission, $t_{peak}$, is associated with the deceleration time of 
the outflow $t_{dec}$, and with the optical emission at $t > t_{dec}$,  
most likely dominated by the forward shock with 
${\mathcal E}_{opt} > {\mathcal E}_m$. This conclusion is similar to the 
results we obtained by considering the early X-ray and optical emissions 
from the tail emission in the prompt shell without evoking a second 
component and a separate origin for the optical and X-ray emissions. In fact, 
using the observed power-law spectrum of the early X-ray, we can estimate 
the expected optical flux by extending the spectrum to the optical bands. 
What we find is summarized in Table \ref{TABLE:optexpect}. The observed 
magnitudes in this table are corrected for the Milky Way 
absorption which is only $\sim 0.1$ magnitude. They are not corrected for 
the host absorption because we do not have any information about the dust 
to gas ratio in the host galaxy. As discussed in Sec.\ref{SECTION:specvar}, 
the host of the most GRBs have values similar to SMC, $A_V \sim 1$ magnitude. 
Although the difference between the expected and the observed may 
partially be due to the absorption in the host, we also notice that 
discrepancies are time dependent, see Table\ref{TABLE:optexpect}. The value 
of $V_\mathrm{obs} - V_\mathrm{pred}$ varies from $\sim 1.5$~mag up to 
$\sim 5$~mag depending on the time interval. The observational errors of the 
observed magnitudes are at most $\pm 0.25$ at $1\sigma$ or $\pm 0.5$ at 
$2\sigma$. They can not explain the large variations as well as differences 
between expectations and observations. Therefore, we conclude that at least 
part of the difference between observations and expectations is intrinsic 
to the fireball. If the early X-ray and optical radiations have different 
origin, then the discrepancies between expectations and observations become 
even larger. This is because we must add the contribution in the optical 
flux of the other component to the contribution of the component producing 
the X-ray. 

We now consider the 
possibility of a reverse shock as the source of the early optical emission. 
The rising slope of the optical light curve at $t < t_{peak}$ is $\sim 2.3$, 
which - in the context of the standard reverse shock model that predicts a 
slope of $\sim 1/3$ - is too steep for the passage of ${\mathcal E}_m$ 
through ${\mathcal E}_{opt}$. The rising slope is a probe of the 
strength of the reverse shock and the observed slope suggests that the reverse 
shock is weak - the dimensionless parameter 
$\zeta \equiv ([3E/(4\pi nm_pc^2)]^{1/3}/\Delta r_0)^{1/2}\Gamma^{-4/3}$ 
\citep{NP2004} presenting the strength of the reverse shock is large 
$\gtrsim 2$. This lower limit on $\zeta$ is valid both if the emission at 
$t < t_{peak}$ is dominated by the forward external shock or if it is from 
a non-relativistic or at most mildly-relativistic reverse shock 
\citep{NP2004}. Therefore, it seems that for this burst the contributions 
of these components are small. Nonetheless, if the early optical emission 
is due to a reverse shock, the temporal slope of the rising phase also 
probes the external density profile. A wind-like density profile would 
produce a significantly slower rise than a uniform external density. The steep 
rise seems to favor a uniform external density over a wind-like density 
profile below and around the deceleration radius $r_{dec}$. More 
specifically, a wind-like external density profile could not produce such 
a sharp rise either by a forward or through a reverse shock even if the 
reverse shock is extremely weak ($\zeta \gg 1$) and the outflow acts as a 
perfect piston as far as the external medium is concerned.

\begin{table}
\caption {Expected and observed optical
magnitudes in the $V$ band\label{TABLE:optexpect}}
\begin{center}
\begin{tabular}{rrrc}
\hline\\
Time + $T_0$~(sec) & $V_\mathrm{pred}$ & $V_\mathrm{obs}$ & 
$\Delta V = V_\mathrm{obs} - V_\mathrm{pred}$ \\
\hline \\
      97 & 10.7 & 15.70 & 5.00 \\
     175 & 13.2 & 14.75 & 1.55 \\
     262 & 11.8 & 15.15 & 3.35 \\
\hline
\end{tabular}
\end{center}
\end{table}

Another possible explanation is that the optical emission is mainly produced 
by the combination of reverse shock and energy injection. The latter is 
one of the possible causes for the slow decay of the optical light curve 
after the peak $\sim t^{-1.3}$ compared to the $\sim t^{-2}$ expected for 
an idealized reverse shock where the outflow has a sharp lower cutoff in 
$\gamma$. The observation of X-ray flares for this burst is also consistent 
with the additional outflow after the prompt shell.

There is a minor break in the optical light (seen more clearly in the data of 
\citep{MVM2006} at $\sim T_0 + 103$~sec, roughly at the end of the bright 
X-ray flares and before the small X-ray/optical flare. This may be the 
evidence of the break in the reverse shock after a relatively flat regime 
following the initial rise \citep{ZM2002}. However, for this argument to 
be viable, the reverse shock should be significantly brighter than the 
forward shock. Regarding the arguments above about the weakness of the 
reverse shock, it does not seem that we can associate this break 
to the reverse shock.

We should also mention that the above break can be also explained as a 
hardening of the radiation at the arrival of a new shell responsible for 
the weak flare in XRT data (see Fig. \ref{FIGURE:xrt_lc}) at 
$\sim T_0 + 1260$~sec. A similar break is also visible in the UVOT light 
curve during the flare at $\sim T_0 + 260$~sec, followed by a slightly 
steeper fall of the optical light curve most probably due to energy 
injection and hardening of the emission (see Fig. \ref{FIGURE:lightcurves}). 
Therefore, considering all the arguments together, the evidence for a dominant 
reverse shock is weak.

\section{Conclusion} \label{SECTION:conclu} 
After years of investigation and especially with the multi-band data 
from {\sl Swift\/} of more than 200 bursts, it is clear that complex physical 
processes are involved not only in the physics of the progenitor but also 
in the collisions between ejected material and the surroundings that at 
first sight seem simple.

In this work we applied the internal/external shock model as the source of 
observed emission to GRB 060607A, one of the most peculiar gamma-ray bursts 
observed by {\sl Swift\/}. We tried to understand and interpret the acquired 
data and related processes partly qualitatively, partly quantitatively. We 
estimated the distance to the central engine where the collisions between 
shells occurred to be $\sim 10^{12}$~cm. This is consistent with the original 
suggestion of the model in which, internal shocks happen at a distance of 
$r \sim 10^{11} - 10^{13}$~cm from the central engine. We also found that 
the external shocks influence the afterglow of 
this burst only at late times; most probably a few thousand seconds after 
the prompt emission. If our interpretations are correct, it seems 
that the coherent magnetic field in the fireball can have relatively long 
lifetime and the energy distribution of electrons can be much more complex 
than a simple power-law. Based on the hardness ratio in the X-ray band we 
showed that there is a separation between population/sources of electrons 
responsible for the late X-ray and optical emission. In particular we 
conjectured the presence of a non-synchrotron, close to thermal component 
in the optical emission from the shocked heated material at late times in 
the afterglow of GRBs. The measurement of the late time optical spectrum 
can verify this hypothesis. A better understanding of the relativistic 
plasma physics is also necessary and will provide realistic models for 
the evolution of fields and charged particles acceleration and energy 
distribution during the shock and their relation with the state of matter 
in the ejecta (fireball).

\appendix
\section{} \label{appendixa}
In this appendix we briefly review the main aspects and results of the 
shock model used for comparison with data. It is a reformulation of the 
relativistic shock models with only one synchrotron emitting shocked layer 
but more detailed parametrization of the physical processes and their time 
variation (Ziaeepour, in prep.). 

We consider a radiative shock between two relativistic spherical shells. 
For an observer in the rest frame of the fast shell, the kinetic energy of 
the falling particles from the slow shell is instantaneously emitted as the 
shells join and merge together. A far observer sees that the fast shell 
absorbs the slow one, its density and/or size increases and its Lorentz factor 
decreases. During coalescence, we distinguish one region of unshocked 
material in each shell, and one shocked zone in which charged particles are 
accelerated by the coherent electric field produced be the shock. In 
presence of a coherent magnetic field, these charged particles irradiate 
their energy as synchrotron radiation. The magnetic field is usually 
considered to be produced by the shock as well \citep{Y1994,WA2004}. It is 
also possible that there is an additional ambient magnetic field produced 
by the central engine. Here we neglect this possibility. This model is a 
simplified version of the relativistic shock model \citep{S1996} 
that considers one shocked region in each side of the shock discontinuity - 
forward and reverse. The lack of a clear evidence for a reverse shock 
emission in any burst, specially during the prompt emission, means that the 
reverse shock in GRBs is weak and the assumption of just one shocked region 
is a good approximation.

Energy-momentum conservation equations in the slow shell/ISM frame is:
\bea
&& \gamma (r^2 \frac{d(n\Delta r)}{dr} + 2r (n\Delta r)) + r^2 (n\Delta r) 
\frac{d\gamma}{dr} = \nonumber \\
&&\quad \quad n_0 \gamma r^2 + \frac{dE_{sy}}{4\pi m c^2dr} 
\label {EQUATION:enercons} \\
&& \beta \gamma (r^2 \frac{d(n\Delta r)}{dr} + 2r (n\Delta r)) + r^2 (n\Delta r) 
d(\beta \gamma) = \nonumber \\
&&\quad \quad n_0 \beta \gamma r^2 + \frac{dE_{sy}}{4\pi m c^2dr} 
\label {EQUATION:momcons}
\eea
where $r$ is the distance from the central engine, $n$ is the number density 
of the fast shell measured in the slow shell frame, $n_0$ is the number 
density of the slow shell in its rest frame, $\Delta r$ is the thickness of 
the shocked synchrotron emitting region, $\gamma$ is the Lorentz factor of 
the fast shell in the slow shell frame and 
$\beta = \sqrt {\gamma^2 - 1} / \gamma$, $m = m_p+m_e \approx m_p$, 
$E_{sy}$ is the total emitted energy, and $c$ is the speed of light. In the 
absence of knowledge about the evolution of microphysics of the shocked 
region, we can not evaluate the evolution of $\Delta r$. However, the 
left hand side of equations \ref{EQUATION:enercons} and \ref{EQUATION:momcons} 
depends on $n\Delta r$ - the column density of shocked region. 

The total power of synchrotron emission depends on the number of emitters. 
If only falling electrons are accelerated, the number of emitting sources 
are proportional to $n_0$. In this case we have to consider a model for the 
evolution of the shocked region. A minimal 
estimation is $\Delta r = \beta \Delta r_0 / \gamma$ where $\Delta r_0$ is a 
constant thickness scale. The physical reason for this choice is that in 
the two extreme cases when the relative Lorentz factor is very large or very 
small, one expects a thin shocked region. In the first case the reason is 
that due  to the large density difference between shells only a thin layer 
of the fast shell is affected (shocked) be the falling particles from slow 
shell. In the second case the shock is soft, the two shells merge smoothly 
and one expects that the size of the turbulent shocked particles be very 
restricted. Another plausible model is the acceleration of all the 
electrons in the shocked region following the formation of a coherent 
electric field in the shock. In this case, the column density of emitters 
is proportional to $n\Delta r$, and therefore an adhoc expression for the 
evolution of $\Delta r$ is not needed. Both these cases are based on the 
assumptions that only a detail knowledge about the micro-physics of the 
matter state and shock can confirm or role out. Nonetheless, if the rest frame 
densities of the shells are close to each other, these approximations 
should give similar results. For simplicity, here we will consider the 
first case which makes the right hand side of \ref{EQUATION:enercons} and 
\ref{EQUATION:momcons} independent of $n$. 

We expect that $\Delta r/r \ll 1$ although in some circumstances the 
thickness of the shells can be comparable with $r$. If 
$r \rightarrow \infty$, the shells become planar. If the shells are not 
spherical but collimated, the formulation of the shock and synchrotron 
radiation is more complex. However, it can be shown that at 
first approximation, the total flux can be considered as proportional to the 
opening angle of the jet. Hence for simplicity we do not discuss the effect of 
the collimation further here.

Variation of the synchrotron energy can be related to the emission power 
$dE_{sy}/dr = dE_{sy}/\beta dt = P_{sy}/\beta$, where $P_{sy}$ is the total 
power of the synchrotron emission. Assuming a power-law distribution for the 
Lorentz factor of the accelerated electrons, $dE_{sy}/dr$ can be written as 
a function of the relative Lorentz factor 
$\gamma$, $\epsilon_e$, $\epsilon_B$, and $q$\footnote{The more popular 
parameter $p$ the energy distribution index of the accelerated electrons is 
related to $q$: $p = q - 1$.}, respectively the fraction of kinetic energy 
of the shell transferred to accelerated electrons, the fraction of kinetic 
energy transferred to a coherent magnetic field, and the number distribution 
index of accelerated electrons \citep {S1998}. We assume that $q$ is a 
constant but $\epsilon_e$, $\epsilon_B$ have power-law dependence on $r$ 
with indices $\alpha_e$ and $\alpha_B$ respectively. Then we can solve 
equations \ref{EQUATION:enercons} and \ref{EQUATION:momcons} to 
find the evolution of $\gamma$ and the column density of synchrotron 
emitting layer $n\Delta r$. These equations are coupled and therefore some 
approximations are necessary to decouple and solve them. In addition, their 
solutions depend on the relative strength of the various parameters. 

At lowest order in $\epsilon$, and when $r/r_0 - 1 \ll 1$, the evolution of 
$\gamma$ can be expressed in the following form:
\be
\frac{\gamma^2 (r)}{\gamma^2 (r_0)} = \frac {1}{1 + F \epsilon + \ldots } 
\quad,\quad \frac {r}{r_0} \equiv 1 + \epsilon, \quad \epsilon \ll 1
\label{EQUATION:defeps}  
\ee
The coefficient $F$ depends on various physical quantities such as initial 
relative Lorentz factor of the colliding shells, their initial densities, 
$\epsilon_e$, $\epsilon_B$, the number distribution index of accelerated 
electrons $q$, and the variation of these quantities with time. They are 
degenerate and the extraction of all of them from available data is not 
possible.

The column density of synchrotron emitting layer at the lowest order and with 
approximations mentioned above is:
\be
\frac{n\Delta r}{n (r_0)\Delta r_0} = \biggl (\frac{r_0^2}{r^2} \biggr ) 
(1 + F_1 \biggl (\frac{r_0^2}{r^2} \biggr )^{\alpha'} + \ldots) 
\label{EQUATION:dens}  
\ee
where the term $r_0^2 / r^2$ is simply due to the adiabatic expansion of 
the shell. The constant $F_1$ depends on various parameters mentioned about. 
The exponent $\alpha'$ consists of 2 parts: a constant and a part dependent on 
$\eta \equiv 2\alpha_e + \alpha_B$. This shows the importance of the 
evolution of coherent electric and magnetic fields in the GRBs. 

Evidently, the only observable outcome of the shell collision for us is the 
synchrotron emission. Its intensity for one charged particle is well known 
\citep {J1999}, and in order to obtain the total intensity we should 
integrate it over the emitting volume and the spectrum of the synchrotron 
emitting charged particles (electrons):
\be
\frac{dI}{d\nu} = 2\sqrt{3}\pi \frac{e^2}{c} V(r) \gamma \int_{\Gamma}^{\infty} 
d\gamma_e n_e (\gamma_e) \frac{\nu}{\nu_c} \int_{\frac{\nu}{\nu_c}}^{\infty}
K_{\frac{5}{3}} (x)dx \label{EQUATION:synchintens}
\ee
where $V(r)$ is the volume of the synchrotron emitting layer, $\nu_c$ is 
synchrotron characteristic frequency, $n_e (\gamma_e)$ is the electron 
density with Lorentz factor $\gamma_e$, $\Gamma$ is the minimum Lorentz 
factor of the accelerated electrons, and finally $K_{\frac{5}{3}}$ is the 
Bessel function. Although $\frac{dI}{d\nu}$ is the main observable quantity, 
it depends on too many unknown parameters and in practice it can not be 
efficiently used. Two other quantities derived from intensity are more useful. 
One is the hardness ratio and the other is the lag of the peak emission 
in different energy bands:
\bea
HR_{ij} & \equiv & \frac{\bar{\nu_i} \frac{dI}{d\bar{\nu_i}} \log \biggl (
\frac{\nu_i^{max}}{\nu_i^{min}} \biggr )}{\bar{\nu_j} \frac{dI}{d\bar{\nu_j}} \log \biggl ( \frac{\nu_j^{max}}{\nu_j^{min}} \biggr )} 
\label{EQUATION:hr} \\
\Delta \epsilon_{ij} & = & \epsilon_i^{peak} - \epsilon_j^{peak} \quad , \quad 
\frac{d}{d\epsilon}\biggl ( \frac{dI}{d\nu} \biggr ) \biggl |_{\epsilon^{peak}} = 0 
\label{EQUATION:depsilon}
\eea
where $i$ and $j$ are different energy bands with logarithmic mean frequency 
$\bar{\nu}$ and minimum and maximum frequency $\nu^{min}$ and $\nu^{max}$, 
respectively. The variable $\epsilon$ is defined in \ref {EQUATION:defeps}. 
The hardness ratio can be determined directly from the data for all energy 
bands. However, the lags are only measurable for fast varying features such 
gamma-ray peaks and X-ray flares. Nonetheless, when they are measurable 
they give valuable information about the distance to the central engine at 
which, the corresponding events have occurred. Note that $\epsilon$'s in 
\ref {EQUATION:depsilon} are dimensionless relative quantities. 
However, the measured lags are not. Therefore, if the lags in multiple 
bands are available, we can estimate the value of the corresponding 
$\epsilon$'s thereby we can measure the initial distance $r_0$.

It can be shown that $HR_{ij}$ and $\Delta \epsilon_{ij}$ are depend on 
$r$ (or equivalently $t$ or $\epsilon$) through the function 
$C(r,\nu) \equiv \nu/\nu_c' \Gamma^2 =\nu/\nu_m $, where 
$\nu_c' \equiv eB/m_e c$ depends only on the magnetic field. Therefore, 
even without any knowledge about other quantities, the evolution of the 
hardness ratio directly shows the evolution of $\nu_m$. The analytical 
expression for $HR_{ij}$ and specially for $\Delta \epsilon_{ij}$ 
are quite involved. Moreover, the integral over the Bessel function in 
\ref{EQUATION:synchintens} must be calculated numerically. Our experience 
shows that the use of asymptotic values of the Bessel function which permit 
to obtain analytical expressions for these quantities leads to large errors 
and unphysical behaviour of the results. The results of numerical 
calculation for a number of combination of parameters will be reported in 
(Ziaeepour, in prep.) and we have used them for analyzing GRB 060607A data.

\section{}\label{appendixb}
Practically in all gamma-ray bursts the spectrum becomes softer after the 
end of the main prompt spikes. Therefore the spectral index of the gamma-ray 
spectrum should be considered as a slowly time-varying quantity. The total 
X-ray flux $f_{\gamma-X}(t)$ expected is approximately:
\bea
f_{\gamma-X}(t) & \approx & \frac {f_\gamma (t)}{E_{max} - E_{min}}
\int_{E_{min}}^{E_{max}} \biggl (\frac{E_X}{\bar {E_\gamma}} \biggr )^
{\alpha_\gamma (t)} {dE_X} \nonumber \\
& \approx & \frac {f_\gamma (t)}{\alpha_\gamma (t) + 1}
\biggl (\frac{E_{max}}{\bar {E_\gamma}} \biggr )^{\alpha_\gamma (t)} 
\label{fluxgx}
\eea
where $\bar {E_\gamma}$ is the logarithmic mean of gamma-ray energy band, 
$E_{max}$ and $E_{min}$ are maximum and minimum of the X-ray energy band, and 
$\alpha_\gamma (t)$ is the slowly varying spectral index.

In the method explained in Sec. \ref{SECTION:specvar} for the extrapolation of 
BAT spectrum to X-ray band, a simple average of the spectral index, usually 
from the total BAT spectrum and early X-ray spectrum is used as an 
approximation to take into account the time evolution of the spectrum. 
Using the same formulation as in 
equation \ref{fluxgx}, the extrapolated BAT light curve to the XRT band 
$f_{BAT-XRT} (t)$ can be estimated as:
\be
f_{BAT-XRT} (t) \approx \frac {N f_{BAT} (t)}{\bar{\alpha} (t) + 1}
\biggl (\frac{E_{max}}{\bar {E_\gamma}} \biggr )^{\bar{\alpha}} 
\label{fluxbatxrt}
\ee
where $\bar{\alpha}$ is the average of BAT and XRT spectrum indices and $N$ 
is a normalisation factor between the two instruments. In general, the cruder 
approximation of \ref{fluxbatxrt} will not agree with the more precise 
expression \ref{fluxgx} unless $\alpha_\gamma (t)$ have roughly a linear and 
slow evolution with time. If these conditions are not fulfilled and/or the 
prompt and the tail emission in X-ray have different origins, the extrapolated 
light curve \ref{fluxbatxrt} should deviate from the observed one, or a break 
appears when one tries to join simultaneous observations by BAT and XRT. 
This break can not be removed by adjusting the 
normalisation factor $N$ in \ref{fluxbatxrt}. Among 40 \swift~GRBs studied 
in \citep{OB2006} only a few of them seems to need such a break to join the 
extrapolated BAT and the observed XRT light curves: GRB 050315, GRB 050713B, 
and GRB 050915B. In conclusion, \ref{fluxbatxrt} is a good approximation of 
\ref{fluxgx} for most bursts and when it is not, the probability of having 
a smooth and overlapping light curve with the XRT observations is very small.


\section*{Acknowledgments}

This research has made use of data obtained through the High Energy
Astrophysics Science Archive Research Center Online Service, provided
by NASA's Goddard Space Flight Center.




\begin{thebibliography}{99}
\bibitem[\protect\citeauthoryear{Achterberg {\etal}}{2001}]{ACH2001}
     Achterberg, A., Gallant, Y.~A., Kirk, J.~G., Guthmann, A.W.,
     2001, {\MRA}, 328, 393

\bibitem[\protect\citeauthoryear{Amano \& Hoshino}{2007}]{ah2006}
     Amano, T. \& Hoshino, M., 
     2007. {\APJ}, 661, 190
\bibitem[\protect\citeauthoryear{Akerlof {\etal}}{1999}]{ABB1999}
     Akerlof, C., Balsano, R., Barthelemy, S., Bloch, J.,
     Butterworth, P., Casperson, D., Cline, T., Fletcher, S.,
     {\etal},
     1999, {\NAT}, 398, 400
\bibitem[\protect\citeauthoryear{Akerlof {\etal}}{2003}]{AKM2003}
     Akerlof, C.~.W., Kehoe, R.~L., McKay, T.~A., Rykoff, E.~S.,
     Smith, D.~A., Casperson, D.~E.; McGowan, K.~E., Vestrand, W.~T.,
     {\etal},
     2003, {\PAS}, 115, 132
\bibitem[\protect\citeauthoryear{Arnaud}{1996}]{A1996}
     Arnaud, K.~A.,
     Astronom,ical Data Analysis Software and Systems V,
     eds.\ Jacoby, G. and Barnes, J., ASP Conf.\ Ser.\ 101, 1

\bibitem[\protect\citeauthoryear{Barthelmy {\etal}}{2005b}]{bar05}
       Bathelmy, S.D., {\etal},
       2005, {\APJ}, 559, 710 

\bibitem[\protect\citeauthoryear{Barthelmy {\etal}}{2005a}]{B2005}
     Barthelmy, S.D., {\etal}, 
     2005, SSRv, 120, 143
\bibitem[\protect\citeauthoryear{Bednarz \& Ostrowski}{1996}]{BO1996}
     Bednarz, J., \& Ostrowski, M.,
     1996, {\MRA}, 283, 447
\bibitem[\protect\citeauthoryear{Bo{\"e}r}{2001}]{B2001}
     Bo{\"e}r, M.,
     2001, Astron.\ Nachr., 322, 343

\bibitem[\protect\citeauthoryear{Burrows {\etal}}{2005}]{BET2005}
     Burrows, D.N., {\etal},
     2005, SSRv, 120, 165

\bibitem[\protect\citeauthoryear{Butler {\etal}}{2006}]{BUT2006}
     Butler, N.~R., {\etal},
     2006 {\APJ}, 652, 1390

\bibitem[\protect\citeauthoryear{Covino {\etal}}{2004}]{CSS2004}
     Covino, S., Stefanon, M., Sciuto, G., Fernandez-Soto, A.,
     Tosti, G., Zerbi, F.~M., Chincarini, G., Antonelli, L.~A.,
     {\etal},
     2004, Proc.\ SPIE, 5492, 1613
\bibitem[\protect\citeauthoryear{Dai \& Lu}{1998}]{DL1998}
     Dai, Z.~G., \& Lu, T.,
     1998, {\AA}, 333, L87
\bibitem[\protect\citeauthoryear{Dai {\etal}}{2007}]{D2007} Dai, X., {\etal}, 
     2007, {\APJ}, 658, 509 
\bibitem[\protect\citeauthoryear{Fenimore {\etal}}{1995}]{Fen1995} Fenimore, E.E., $_IN0T$ Zand, J.J.M., 
     Norris, J.P., Bonnel, J.T., Nemiroff, R.J., 
     1995, {\APJ}, 101, 448L 
\bibitem[\protect\citeauthoryear{Fenimore {\etal}}{1996}]{Fen1996} Fenimore, E.E., Madras, C.D. \& 
     S. Nayakshin, 
     1996, {\APJ}, 473, 998 
\bibitem[\protect\citeauthoryear{Frail {\etal}}{2001}]{FKS2001}
     Frail, D.~A., Kulkarni, S.~R., Sari, R., Djorgovski, S.~G.,
     Bloom, J.~S., Galama, T.~J., Reichart, D.~E., Berger, E., {\etal},
     2001, {\APJl}, 562, L55     
\bibitem[\protect\citeauthoryear{Fynbo {\etal}}{2001}]{FGD2001}
     Fynbo, J.~P.~U., Gorosabel, J., Dall, T.~H., Hjorth, J.,
     Pedersen, H., Andersen, M.~I., M{\o}ller, P., Holland, S.~T.,
     {\etal},
     2001, {\AA}, 373, 796
\bibitem[\protect\citeauthoryear{Gehrels {\etal}}{2004}]{GCG2004}
     Gehrels, N., Chincarini, G., Giommi, P., Mason, K.~O.,
     Nousek, J.~A., Wells, A.~A., White, N.~E., Barthelmy, S.~D.,
     {\etal},
     2004, {\APJ}, 611, 1005

\bibitem[\protect\citeauthoryear{Goad {\etal}}{2007a}]{GOA2007}
     Goad, M., Page, K.~L., Godet, O., Beardmore, A., Osborne, J., {\etal},
     2007, {\AA}, 468, 103

\bibitem[\protect\citeauthoryear{Goad {\etal}}{2007b}]{G2007}
     Goad, M., {\etal}, submitted, arXiv:0708.0986
\bibitem[\protect\citeauthoryear{Hjorth {\etal}}{2003}]{HMG2003}
     Hjorth, J, M{\o}ller, P., Gorosabel, J., Fynbo, J.~P.~U.,
     Toft, S., Jaunsen, A.~O., Kaas, A.~A., Pursimo, T.,
     {\etal},
     2003, {\APJ}, 597, 699
\bibitem[\protect\citeauthoryear{Holland {\etal}}{2007}]{HBG2007}
     Holland, S.~T., Boyd, P., Gorosabel, J., Hjorth, J.,
     Schady, P., Thomsen, B., Augusteijn, T., Blustin, A.~J.,
     {\etal},
     2007, {\AST}, 133, 12
\bibitem[\protect\citeauthoryear{Ioka {\etal}}{2006}]{ITY2006}
     Ioka, K., Toma, K., Yamazaki, R., \& Nakamura, T.,
     2006, {\AA}, 458, 71

\bibitem[\protect\citeauthoryear{Jackson}{1999}]{J1999}
J.D. Jackson, ``Classical Electrodynamics'', 1999, 3$^{rd}$ ed., Wiley, Inc.

\bibitem[\protect\citeauthoryear{Jensen {\etal}}{2001}]{JFG2001}
     Jensen, B.~L., Fynbo, J.~P.~U., Gorosabel, J., Hjorth, J.,
     Holland, S.~T., M{\o}ller, P., Thomsen, B., Bj{\"o}rnsson, G.,
     2001, {\AA}, 370, 909
\bibitem[\protect\citeauthoryear{Kahn {\etal}}{2006}]{KKZ2006}
     Kahn, D.~A., Klose, S., \& Zeh, A.,
     2006, {\APJ}, 641, 993

\bibitem[\protect\citeauthoryear{Kirk {\etal}}{2000}]{KI2000}
     Kirk, J.~G., Guthmann, A.W., Gallant, Y.~A., Achterberg, A.,
     2000, {\APJ}, 542, 235

\bibitem[\protect\citeauthoryear{Kumar {\etal}}{2006}]{KMP2006}
     Kumar, P., McMahon, E., Panaitescu, A., {\etal},
     2006, in prep.
\bibitem[\protect\citeauthoryear{Ledoux {\etal}}{2006}]{LVS2006}
     Ledoux, C., Vreeswijk, P., Smette, A., Jaunsen, A.,
     \& Kaufer, A.,
     2006, GCN Circ.\ 5237
\bibitem[\protect\citeauthoryear{Liang {\etal}}{2006}]{LZO2006}
     Liang, E.~W., Zhang, B., O'Brien, P.~T., Willingale, R.,
     Angelini, L., Burrows, D.~N., Campana, S., Chincarini,G.,
     {\etal},
     2006, {\APJ}, 646, 351

\bibitem[\protect\citeauthoryear{Misra {\etal}}{2007}]{M2007}
     Misra, K., {\etal},
     2007, {\AA}, 464, 903

\bibitem[\protect\citeauthoryear{Molinari {\etal}}{2007}]{MVM2006}
     Molinari, E., Vergani, S.~D., Malesani, D., Covino, S.,
     D'Avanzo, P., Chincarini, G., Zerbi, F.~M., Antonelli, L.~A.,
     {\etal},
     2006, astro-ph/0612607
\bibitem[\protect\citeauthoryear{Nakar \& Piran}{2002}]{NP2002} Nakar, E., \& Piran, T., 2002, {\MRA}, 331, 40 
\bibitem[\protect\citeauthoryear{Nakar \& Piran}{2004}]{NP2004}
     Nakar, E., \& Piran, T.,
     2004, {\MRA}, 353, 647
\bibitem[\protect\citeauthoryear{Nousek {\etal}}{2006}]{NKG2006}
     Nousek, J.~A., Kouveliotou, C., Grupe, D., Page, K.~L.,
     Granot, J., Ram{\'\i}rez-Ru{\'\i}z,, E., Patel, S.~K., Burrows, D.~N.,
     {\etal},
     2006, {\APJ}, 642, 389
\bibitem[\protect\citeauthoryear{Nysewander \& Haislip}{2006}]{NH2006}
     Nysewander, M., \& Haislip. J.,
     2006, GCN Circ.\ 5236

\bibitem[\protect\citeauthoryear{Nysewander {\etal}}{2007}]{NH2007}
     Nysewander, M., {\etal},
     2007, submitted, arXiv:0708.3444

\bibitem[\protect\citeauthoryear{Oates {\etal}}{2006}]{O2006}
     Oates, S.R., {\etal},
     2006, GCN Circ.\  5243

\bibitem[\protect\citeauthoryear{O'Brien {\etal}}{2006}]{OB2006}
     O'Brien, P.~T., {\etal},
     2006, {\APJ}, 647, 1213

\bibitem[\protect\citeauthoryear{Page {\etal}}{2006}]{P2006}
     Page, K, Goad, M. \& Beardmore, A.,
     2006, GCN Circ.\ 5240
\bibitem[\protect\citeauthoryear{Page {\etal}}{2007}]{P2007}
     Page, K.~L., {\etal},
     2007, {\APJ}, 663, 1125
\bibitem[\protect\citeauthoryear{Pei}{1992}]{P1992}
     Pei, Y.,
     1992, {\APJ}, 395, 130
\bibitem[\protect\citeauthoryear{Predehl \& Schmitt}{1995}]{PS1995}
     Predehl, P., \& Schmitt, J.~H.~M.~M.,
     1995, {\AA}, 293, 889
\bibitem[\protect\citeauthoryear{Perna {\etal}}{2003}]{PKG2003}
     Perna, R., Lazzati, D., \& Fiore, F.,
     2003, {\APJ}, 585, 775
\bibitem[\protect\citeauthoryear{P{\'e}rez-Ram{\'\i}rez {\etal}}{2004}]{PMR2001}
     P{\'e}rez-Ram{\'\i}rez, D., Merlioni, A., \& Rees, M.~J.,
     2004, Astron.\ Nachr., 324, 1147
\bibitem [\protect\citeauthoryear{Ramirez-Ruiz \& Merloni}{2001}]{RM2001}Ramirez-Ruiz, E.F. \& 
     Merloni, A., 
     2001, {\MRA}, 320, L25 

\bibitem [\protect\citeauthoryear{Ramirez-Ruiz {\etal}}{2001}]{RET2001}Ramirez-Ruiz, E.F, Merloni, A., 
     Rees, M.~J.,
     2001, {\MRA}, 324, 1147

\bibitem [\protect\citeauthoryear{Roming {\etal}}{2005}]{RO2005}
     Roming, P.W.A., {\etal},
     2005, SSRv, 120, 95

\bibitem[\protect\citeauthoryear{Rees \& M{\'e}sz{\'a}ros}{1994}]{RM1994}
     Rees, M.~J., \&  M{\'e}sz{\'a}ros, P.,
     1994, {\APJl}, 430, L93
\bibitem[\protect\citeauthoryear{Rees \& M{\'e}sz{\'a}ros}{1998}]{RM1998}
     Rees, M.~J., \&  M{\'e}sz{\'a}ros, P.,
     1998, {\APJl}, 496, L1
\bibitem[\protect\citeauthoryear{Rees \& M{\'e}sz{\'a}ros}{2000}]{RM2000}
     Rees, M.~J., \&  M{\'e}sz{\'a}ros, P.,
     2000, {\APJl}, 545, L73
\bibitem[\protect\citeauthoryear{Reville {\etal}}{2006}]{RKD2006}
     Reville, B., Kirk, J.~G., \&  Duffy, P.
     2006, Plasma Phys.\ and Controlled Fusion, 48, 1741
\bibitem[\protect\citeauthoryear{Rhoads}{1999}]{R1999}
     Rhoads, J.~E.,
     1999, {\APJ}, 525,737
\bibitem[\protect\citeauthoryear{Rieger {\etal}}{2006}]{RBD2006}
     Rieger, F.~M., Bosch-Ramon, V., \&  Duffy, P.,
     2007, {\APS}, in press, astro-ph/0610141
\bibitem[\protect\citeauthoryear{Romano {\etal}}{2006}]{RCC2006}
     Romano, P., Campana, S., Chincarini, G., Cummings, J.,
     Cusumano, G., Holland, S.~T., Mangano, V., Mineo, T.,
     {\etal},
     2006, {\AA}, 456, 917 
\bibitem [\protect\citeauthoryear{Sari {\etal}}{1996}]{S1996}Sari, R., Narayan, R., Piran, T.,
     1996, {\APJ}, 473, 204  

\bibitem [\protect\citeauthoryear{Sari}{1997}]{S1997}
     Sari, R., 1997, {\APJ}, 489, L37
\bibitem [\protect\citeauthoryear{Sari {\etal}}{1998}]{S1998}Sari, R., Piran, T., Narayan, R., 
     1998, {\APJl}, 497, L17 
\bibitem[\protect\citeauthoryear{Sari {\etal}}{1999}]{SPH1999}
     Sari, R., Piran, T., \& Halpern, J.~P.,
     1999, {\APJl}, 519, L17
\bibitem[\protect\citeauthoryear{Sari \& M{\'e}sz{\'a}ros}{2000}]{SM2000}
     Sari, R., \&  M{\'e}sz{\'a}ros, P.,
     2000, {\APJl}, 535, L33
\bibitem[\protect\citeauthoryear{Savitzky \& Golay}{1964}]{SG1964} Savitzky, A., \& Golay, M.J.E, 1964, 
     Analytical Chemistry, 36, 1627 
\bibitem[\protect\citeauthoryear{Schady {\etal}}{2006}]{S2006} 
     Schady, p., {\etal}, {\MRA}, in press, astro-ph/0611089 

\bibitem[\protect\citeauthoryear{Schady {\etal}}{2007}]{S2007} 
     Schady, p., {\etal}, 
     2007, {\MRA}, 377, 273

\bibitem[\protect\citeauthoryear{Schlegel {\etal}}{1998}]{SFD1998}
     Schlegel, D.~J., Finkbeiner, D.~P., \& Davis, M.,
     1998, {\APJ}, 500, 525 


\bibitem[\protect\citeauthoryear{Shen {\etal}}{2006}]{SHE2006}
      Shen, R., Kumar, P., \& Robinson, E.~L.,
      2006, {\MRA}, 371, 1441

\bibitem[\protect\citeauthoryear{Starling {\etal}}{2007a}]{STG2007}
     Starling, R.~L.~C., {\etal}, 
     2007, GCN Circ.\ 6542

\bibitem[\protect\citeauthoryear{Starling {\etal}}{2007b}]{ST2007}
     Starling, R.~L.~C., {\etal}, 
     2007, submitted to {\MRA}

\bibitem[\protect\citeauthoryear{Stamatikos {\etal}}{2007}]{STAM2007}
     Stamatikos, M., {\etal},
     2007, GCN Rep.\ 25.1

\bibitem[\protect\citeauthoryear{Stratta {\etal}}{2004}]{SFA2004}
     Stratta, G., Fiore, F., Antonelli, L.~A., \& De~Pasquale, M.,
     2004, {\APJ}, 608, 846

\bibitem[\protect\citeauthoryear{Tueller {\etal}}{2006}]{TBB2006}
     Tueller, J., Barbier, L., Barthelmy, S., Cummings, J.,
     Fenimore, E., Gehrels, N., Hullinger, D., Koss, M.,
     {\etal},
     2006, GCN Circ.\ 5242
\bibitem[\protect\citeauthoryear{Vestrand {\etal}}{2002}]{VBB2002}
     Vestrand, W~.T. Borozdin, K.~N., Brumby, S.~P., Casperson, D.~E.,
     Fenimore, E.~E., Galassi, M.~C., McGowan, K., Perkins, S.~J.,
     {\etal}
     2002, Proc.\ SPIE, 4845, 126
\bibitem[\protect\citeauthoryear{Vestrand {\etal}}{2005}]{VWW2005}
     Vestrand, W.~T., Wo{\'z}niak, P.~R., Wren, J.~A., Fenimore, E.~E.,
     Sakamoto, T., White, R.~R., Casperson, D., Davis, H.,
     {\etal},
     2005, {\NAT}, 435, 178
\bibitem[\protect\citeauthoryear{Vestrand {\etal}}{2006}]{VWW2006}
     Vestrand, W.~T., Wren, J.~A., Wo{\'z}niak, P.~R., Aptekar, R.,
     Golentskii, S., Pal'shin, V., Sakamoto, T., White, R.~R.,
     {\etal},
     2005, {\NAT}, 442, 172

\bibitem[\protect\citeauthoryear{Watson {\etal}}{2007}]{W2007}
     Watson, D., {\etal},
     2007, {\APJ}, 660, 101

\bibitem[\protect\citeauthoryear{Waxman \& Draine}{2000}]{WD2000}
     Waxman, E., \& Draine, B.~T.
     2000, {\APJ}, 537, 796

\bibitem[\protect\citeauthoryear{Wei \& Lu}{2002}]{wl2002}
     Wei, D.M. \& Lu, T., 
     2002, {\MRA}, 332, 994
\bibitem[\protect\citeauthoryear{Wiersma \& Achterberg}{2004}]{WA2004}
     Wiersma, J., \& Achterberg, A.,
     2004, {\AA}, 428, 365

\bibitem[\protect\citeauthoryear{Willingale {\etal}}{2007}]{WI2007}
     Willingale, R., {\etal},
     2007, {\APJ}, 662, 1093
\bibitem[\protect\citeauthoryear{Yang {\etal}}{1994}]{Y1994}
     Yang, T.~Y.~B., {\etal},
     1994, Phys.\ Plasma, 1, 3059

\bibitem[\protect\citeauthoryear{Yost {\etal}}{2007}]{Y2007}
     Yost, S.~A., {\etal},
     2007, {\APJ}, 657, 925
\bibitem[\protect\citeauthoryear{Zhang \& M{\'e}sz{\'a}ros}{2002}]{ZM2002}
     Zhang, B., \&  M\'es\'zaros, P.,
     2002, {\APJ}, 566, 712
\bibitem[\protect\citeauthoryear{Zhang \& Kobayashi}{2005}]{ZK2005}
     Zhang, B., \& Kobayashi, S.,
     2005, {\APJ}, 628, 315
\bibitem[\protect\citeauthoryear{Zhang {\etal}}{2006}]{ZFD2006}
     Zhang, B., Fan, Y.~Z., Dyks, J., Kobayashi, S.,
     M{\'e}sz{\'a}ros, P., Burrows, D.~N., Nousek, J.~A., \& Gehrels, N.,
     2006, {\APJ}, 642, 354 

\bibitem[\protect\citeauthoryear{Zhang {\etal}}{2007a}]{ZLA2007}
     Zhang, B.~B., Liang,E.~N., Zhang, B.,
     2007, {\APJ}, 666, 1002

\bibitem[\protect\citeauthoryear{Zhang {\etal}}{2007b}]{ZLB2007}
     Zhang, B.~B., Liang,E.~N., Zhang, B.,
     2007, {\APJ}, in press, arXiv:0705.1373

\bibitem[\protect\citeauthoryear{Ziaeepour {\etal}}{2006}]{HZGCN}
     Ziaeepour, H., {\etal},
     2006, GCN Circ.\  5233
\end{thebibliography}
\end{document}